\documentclass[10pt, journal]{IEEEtran}
\IEEEoverridecommandlockouts
\usepackage{cite}
\usepackage{hyperref}
\usepackage{amsmath,amssymb,amsfonts}
\usepackage{algorithmic}
\usepackage{graphicx}
\usepackage{textcomp}
\usepackage{xcolor}
\usepackage{multirow}
\usepackage{wrapfig}
\usepackage{enumitem}
\usepackage{svg}
\usepackage{scalerel}
\usepackage{tikz}
\usepackage{pifont} 
\usetikzlibrary{svg.path}
\usepackage{booktabs}
\usepackage{multirow}
\usepackage{array}
\definecolor{orcidlogocol}{HTML}{A6CE39}
\tikzset{
  orcidlogo/.pic={
    \fill[orcidlogocol] svg{M256,128c0,70.7-57.3,128-128,128C57.3,256,0,198.7,0,128C0,57.3,57.3,0,128,0C198.7,0,256,57.3,256,128z};
    \fill[white] svg{M86.3,186.2H70.9V79.1h15.4v48.4V186.2z}
                 svg{M108.9,79.1h41.6c39.6,0,57,28.3,57,53.6c0,27.5-21.5,53.6-56.8,53.6h-41.8V79.1z M124.3,172.4h24.5c34.9,0,42.9-26.5,42.9-39.7c0-21.5-13.7-39.7-43.7-39.7h-23.7V172.4z}
                 svg{M88.7,56.8c0,5.5-4.5,10.1-10.1,10.1c-5.6,0-10.1-4.6-10.1-10.1c0-5.6,4.5-10.1,10.1-10.1C84.2,46.7,88.7,51.3,88.7,56.8z};
  }
}

\newcommand\orcidicon[1]{\href{https://orcid.org/#1}{\mbox{\scalerel*{
\begin{tikzpicture}[yscale=-1,transform shape]
\pic{orcidlogo};
\end{tikzpicture}
}{|}}}}
\usepackage{hyperref} 

\def\BibTeX{{\rm B\kern-.05em{\sc i\kern-.025em b}\kern-.08em
		T\kern-.1667em\lower.7ex\hbox{E}\kern-.125emX}}

\usepackage{titlesec}

\begin{document}
	

\title{A Comprehensive Survey on the Security of Smart Grid: Challenges,  Mitigations, and Future\\ Research Opportunities}

\author{Arastoo~Zibaeirad$^{\orcidicon{0009-0003-4213-9940}}$, Farnoosh~Koleini$^{\orcidicon{0000-0002-4650-7427}}$, Shengping Bi$^{\orcidicon{0009-0006-2738-224X}}$, Tao Hou$^{\orcidicon{0000-0003-3775-6170}}$,~\IEEEmembership{Member,~IEEE}, Tao~Wang\IEEEauthorrefmark{1}$^{\orcidicon{0000-0001-9744-107X}}$,~\IEEEmembership{Member,~IEEE} 
\thanks{\IEEEauthorrefmark{1}Corresponding Author}
\thanks{Arastoo~Zibaeirad and Farnoosh~Koleini are with the Department of Computer Science, University of North Carolina at Charlotte, Charlotte, NC 28223 USA (e-mail: azibaeir@uncc.edu; fkoleini@uncc.edu)}
\thanks{Shengping Bi, Tao Hou, and Tao Wang are with the Department of Computer Science and Engineering, University of North Texas, Denton, TX 76205 USA (e-mail: shengpingbi@my.unt.edu;  tao.hou@unt.edu; tao@unt.edu)}
}
\maketitle

\begin{abstract}
In this study, we conduct a comprehensive review of smart grid security, exploring system architectures, attack methodologies, defense strategies, and future research opportunities. We provide an in-depth analysis of various attack vectors, focusing on new attack surfaces introduced by advanced components in smart grids. The review particularly includes an extensive analysis of coordinated attacks that incorporate multiple attack strategies and exploit vulnerabilities across various smart grid components to increase their adverse impact, demonstrating the complexity and potential severity of these threats. Following this, we examine innovative detection and mitigation strategies, including game theory, graph theory, blockchain, and machine learning, discussing their advancements in counteracting evolving threats and associated research challenges. In particular, our review covers a thorough examination of widely used machine learning-based mitigation strategies, analyzing their applications and research challenges spanning across supervised, unsupervised, semi-supervised, ensemble, and reinforcement learning. Further, we outline future research directions and explore new techniques and concerns. We first discuss the research opportunities for existing and emerging strategies, and then explore the potential role of new techniques, such as large language models (LLMs), and the emerging threat of adversarial machine learning in the future of smart grid security.

\end{abstract}


		
\section{Introduction}

\IEEEPARstart{C}{onventional} power grids have been integral to electricity distribution for decades, serving as the core infrastructure for energy transmission from power plants to end-users \cite{amin2005toward}. However, the growing electricity demand and increasing integration of renewable energy resources have urged a more advanced and enhanced power grid system. As a result, the smart grid, a sophisticated electrical grid system that incorporates advanced information and communication technologies (ICT), has been introduced to enhance the efficiency, sustainability, and reliability of electricity distribution \cite{fang2011smart}.

\begin{table*}[t]
\renewcommand\arraystretch{1.5}
\centering
\setlength{\tabcolsep}{8pt}
\caption{Comparisons of Our Survey Paper With The Existing Surveys}
\begin{tabular}
{|c|c|c|c|c|c|c|}
  \hline
  \textbf{Key Criteria} & \centering{\textbf{Our Survey}} & \parbox{1cm}{\centering \textbf{\cite{ghiasi2023comprehensive}} \\ (2023)} & \parbox{1cm}{\centering \textbf{\cite{hasan2023review}} \\ (2023)} & \parbox{1cm}{\centering \textbf{\cite{kayan2022cybersecurity}} \\ (2022)} & \parbox{1cm}{\centering \textbf{\cite{faquir2021cybersecurity}} \\ (2021)} & \parbox{1cm}{\centering \textbf{\cite{nguyen2020cyber}} \\ (2020)}\\
  \hline

Comprehensive Review of Smart Grid Layers & \ding{51} & \ding{51} & \ding{51} & \ding{51} & \ding{51} & \ding{115} \\ \hline

Holistic Integration of Cyber, Cyber-Physical, and Physical Layers & \ding{51} & \ding{115} & \ding{115} & \ding{51} & \ding{115} & \ding{55}\\ \hline

Architecture of Smart Grid & \ding{51} & \ding{51} & \ding{51} & \ding{51} & \ding{51} & \ding{115}\\ \hline

Classification of Attacks & \ding{51} & \ding{51} & \ding{51} & \ding{51} & \ding{51} & \ding{55}\\ \hline

Emerging Detection and Mitigation Solutions & \ding{51} & \ding{51} & \ding{115} & \ding{115} & \ding{115} & \ding{55} \\ \hline

Include Game Theory Approaches & \ding{51} & \ding{55} & \ding{55} & \ding{55} & \ding{55} & \ding{55}\\ \hline

Include Graph Theory Approaches & \ding{51} & \ding{55} & \ding{55} & \ding{55} & \ding{55} & \ding{51}\\ \hline

Include Blockchain Approaches & \ding{51} & \ding{51} & \ding{55} & \ding{55} & \ding{55} & \ding{55}\\ \hline

Include ML Approaches & \ding{51} & \ding{115} & \ding{51} & \ding{51} & \ding{115} & \ding{55}\\ \hline

Role of Adversarial Machine Learning & \ding{51} & \ding{55} & \ding{55} & \ding{55} & \ding{55} & \ding{55}\\ \hline

Role of Large Language Models (LLMs) & \ding{51} & \ding{55} & \ding{55} & \ding{55} & \ding{55} & \ding{55}\\ \hline

Discuss Challenges of Each Detection/Mitigation Technique & \ding{51} & \ding{51} & \ding{115} & \ding{115} & \ding{115} & \ding{55}\\ \hline

Discuss Future Research Directions & \ding{51} & \ding{51} & \ding{51} & \ding{51} & \ding{115} & \ding{115}\\ \hline

\end{tabular}
\begin{flushleft}
\centering
\ding{51}: Fully addressed; \ding{115}: Partially addressed; \ding{55}: Not addressed at all.
\end{flushleft}
\label{table:related_work}
\end{table*}

The ICT integration in smart grids enables bidirectional flows of both electricity and information, positioning smart grids as the modern backbone of energy infrastructure \cite{gungor2011smart}. By leveraging ICTs, smart grids facilitate dynamic pricing, effective load management, and improved grid operations, thereby mitigating power outages and enhancing system stability \cite{farhangi2009path}\cite{wang2011survey}. In addition, intelligent entities such as smart meters and advanced distribution management systems (ADMS) enable demand-side management and substantially reduce the operational and management costs by employing advanced fault detection and automation techniques \cite{yan2012survey,dileep2020survey,fang2011smart}. Smart grids also facilitate the integration of a diverse range of energy resources. Particularly, the decentralized power generation via distributed energy resources (DERs), reduces transmission losses and improves the energy efficiency \cite{rahimi2010demand}. The integration of renewable energy systems like solar panels and wind turbines delivers a more sustainable and eco-friendly energy production \cite{ahmad2022data, mahmood2015overview}. The adoption of vehicle-to-grid technology as well as the development of microgrids offers a reliable energy supply in the event of outages or disasters, further enhancing the grid resilience \cite{tan2016integration,yoldacs2017enhancing}. 

Although smart grids offer numerous advantages, the increased connectivity and complexity of smart grids also expose them to various security threats. Smart grids have been identified as potential targets for various attacks, which can exploit vulnerabilities in the ICT infrastructure to disrupt power supply, compromise sensitive data, or inflict physical damage to grid components. For example, the Stuxnet attack, uncovered in 2010, aimed at compromising the industrial control systems to damage the Iranian nuclear facilities \cite{langner2011stuxnet}. In 2015, attackers penetrated the information systems of energy distribution companies in Ukraine, leading to significant power outages. The Dragonfly 2.0 campaign, identified in 2017 has demonstrated its capability to compromise the security of energy sector network in the United States and Europe \cite{symantec2017dragonfly}. These examples highlight the critical need for adequate security measures to protect the smart grid infrastructure. 

Towards this, multiple strategies have been proposed in smart girds to improve their security. Specifically, smart grids incorporate advanced monitoring and detection techniques to provide real-time alerts and responses to potential attacks \cite{ten2010cybersecurity}. The self-healing mechanisms facilitate quick recovery from attacks and minimize adverse impacts \cite{amin2002security}. Data exchange has been protected by encrypted and authenticated communication protocols \cite{yan2012survey}. Smart grids also foster collaboration among stakeholders through shared threat intelligence and best security practices, enhancing the security of the entire grid infrastructure \cite{zhang2011distributed}. While conventional security strategies such as cryptography, authentication, and intrusion detection systems (IDS) are essential to preserve the security of smart grids, they might not be sufficient to defend against the continuously evolving and increasingly sophisticated cyber attacks. The complicated structure and heterogeneous components of smart grids impose a significant challenge in combating advanced cyber attacks, necessitating innovative and efficient solutions. 

To defend against evolving attack threats, emerging techniques such as machine learning, blockchain, graph theory, and game theory have been extensively investigated and have proven their capability in understanding and mitigating smart grid security threats. However, these techniques also face substantial challenges. Adversarial machine learning attacks, in particular, pose a considerable threat to machine learning-based strategies, necessitating ongoing research and development to enhance the resilience of smart grid systems. Recently, large language models (LLMs) have also demonstrated great potential in cybersecurity and attack detection, offering insights into their offensive security capabilities \cite{shao2024empirical}. 

In this study, we aim to conduct a comprehensive literature review to examine different network attacks targeting smart grids and investigate emerging mitigation strategies that show potential in better preventing and detecting advanced threats. We carry out a detailed examination of various attack vectors, detection and mitigation strategies, and emerging threats of adversarial machine learning attacks and LLMs, providing a holistic overview of potential security challenges and solutions within the smart grids.

\subsection{Key Contributions}
The main contributions of this paper can be summarized as follows:

\begin{itemize}

\item We provide a systematic study of smart grid architecture, by defining and categorizing them to three interconnected layers including the physical, cyber-physical, and cyber layers. This holistic view offers a multidimensional understanding of smart grid attacks and defenses. 

\item We conduct a review of the literature on various attacks in smart grids, covering cyber, cyber-physical, and coordinated attacks. Our work includes a detailed classification of these attacks, providing a structured understanding of the threats in smart grids. 
				
\item We critically review and compare emerging detection and mitigation solutions, including game theory, graph theory, blockchain, and machine learning techniques, highlighting their potential effectiveness and possible shortcomings.

\item We explore LLMs and adversarial machine learning, emphasizing their significance and potential impacts on the future of smart grid security.

\end{itemize}

\subsection{Related Works}

Multiple surveys have been conducted, each examining different security perspectives of smart grids.

Ghiasi et al. \cite{ghiasi2023comprehensive} delve into the analysis of different cyber attack models, highlighting their characteristics and applicability to understand the vulnerabilities and threats facing these systems. It also explores cutting-edge security approaches like blockchain and quantum computing, positioning these technologies as innovative solutions for enhancing security and resilience in smart grids. Hasan et al. \cite{hasan2023review} delve into the cyber-physical aspects of smart grids. Additionally, it compiles relevant security standards and protocols, serving as a useful resource for implementing or updating security measures. Otuoze et. al. \cite{otuoze2018smart} shows a detailed classification of smart grid security threats into technical and non-technical sources. Furthermore, the paper differentiates between technical sources, which include infrastructural, operational, and data management security, and non-technical sources like environmental factors and regulatory policies. Peng et al. \cite {peng2019survey} and Faquir et al. \cite{faquir2021cybersecurity} review various attack scenarios including denial of service, false data injection, and others, providing insights into how these attacks affect smart grid operations and suggesting defense strategies to mitigate their impact. Liu et al. \cite{liu2012cyber} offer detailed recommendations for policy and operational changes to enhance the security and privacy of smart grid systems, emphasizing the importance of developing standards and protocols that address the unique requirements of these systems. Krishnan et al. \cite{mololoth2023blockchain} identify and discuss the benefits and challenges of applying blockchain and machine learning (ML) techniques within smart grid applications. Jokar et al. \cite{jokar2016survey} cover the application of cryptography in smart grids, addressing key management challenges and suggesting cryptographic measures to secure communications and protect data.

Unlike previous works that focus predominantly on specific aspects of smart grid security such as cyber attacks or the integration of certain technologies like blockchain and ML, our review paper comprehensively explores all three interconnected layers of the smart grid. This approach allows for a better understanding of how vulnerabilities at one layer can affect others, providing a more integrated perspective on smart grid security. By doing so, we address a critical and rapidly evolving area that poses significant risks to smart grid security. While previous surveys have discussed various traditional and modern security techniques, our paper highlights advanced detection and mitigation strategies that employ cutting-edge technologies such as game theory, graph theory, blockchain, and a comprehensive use of ML techniques. This not only covers a wide range of potential solutions but also discusses their practical implications and effectiveness in real-world scenarios. Also, we explore the potential role of LLMs and adversarial machine learning models in smart grids, a topic that is relatively unexplored in existing literature. This provides readers with practical insights into choosing potential security measures based on specific needs and threat landscapes in the future. In Table \ref{table:related_work}, we summarized our survey paper's key contributions and compared it to five recent survey papers based on several key criteria.

\subsection{Structure of the Paper}

\begin{figure}[htbp]
\centerline{\includegraphics[width=2.5in]{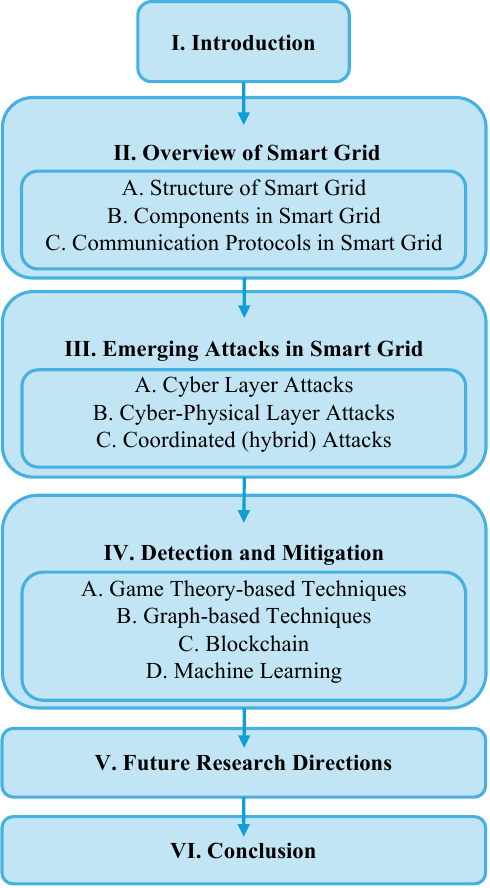}}
\caption{Structure of This Study.}
\label{figure:outline}
\end{figure}

Our survey provides a comprehensive examination of smart grid security, focusing on its structure, emerging attack vectors, detection and mitigation strategies, and future research directions. Figure \ref{figure:outline} visually summarizes the organization and structure of our study.

Section \ref{section:Overview} lays the foundation by detailing the structure, components, and communication protocols of the smart grid. We partition the system into three interconnected layers: the physical layer, the cyber-physical layer, and the cyber layer. The physical layer includes components related to the generation, transmission, distribution, and consumption of electric power. The cyber-physical layer integrates the physical layer with sensing, measurement, and control systems, bridging the gap between physical infrastructure and the digital control system. The cyber layer forms the communication and data exchange backbone, encompassing data communication, networking, supervisory, and management.

Section \ref{section:attack} explores emerging attacks in smart grids. We particularly examine the vulnerabilities associated with new components of smart grids, i.e., cyber and cyber-physical layers. We also investigate coordinated attacks that can affect multiple layers simultaneously, demonstrating the complexity and potential severity of these threats. 

Section \ref{section:detection} focuses on detection and mitigation techniques. As conventional security measures like cryptography, authentication, and network protocols have been intensively investigated, our study shifts towards exploring innovative detection and mitigation approaches including game theory, graph theory, blockchain, and machine learning. These advanced techniques are crucial for future smart grid security, helping to identify and mitigate threats more effectively.

Section \ref{section:future} outlines future research directions to further enhance the smart grid security. We present potential research directions for emerging detection and mitigation techniques, and explore new methods and concerns, particularly LLM applications and adversarial machine learning attacks. LLMs have the potential to understand complex attack patterns and detect sophisticated zero-day attacks, which are crucial to improve smart grid security. Adversarial ML attacks pose significant threats by misleading existing models and bypassing detectors. Therefore, a thorough examination of LLMs and adversarial attacks adds a critical dimension to our study.

Section \ref{section:conclusion} concludes the paper by summarizing the key findings and emphasizing the importance of continued research and development in smart grid security.


\section{Overview of Smart Grid}
\label{section:Overview}

In this section, we provide a brief overview of smart grids from perspectives of its structures, components, and protocols.

\subsection{Structure of Smart Grid}

Smart grids can be built on different architectures given their specific operational concentrations, including management, distribution, communication, security, and scalability. For example, microgrid and peer-to-peer architectures focus on electricity generation and distribution, facilitating local energy production and direct energy trading among participants. Meanwhile, hierarchical and cloud-based architectures concentrate on the scalable and centralized energy management, enhancing control and data analytic capabilities across the smart grid cyber infrastructure. On the other hand, zero trust and end-to-end communication architectures aim at preserving security and ensuring secure, efficient data exchange. While each architectural model focuses on specific aspects, a cohesive, layered framework is essential for the design and functionality of smart grids. Santacana et al. \cite{santacana2010getting} propose a four-layer model comprising power conversion/transport/storage/consumption layer, sensor/actuator layer, communication layer, and decision intelligence layer. Another study \cite{stoustrup2019smart} presents a six-layered framework, namely the physical layer, control layer, data communication layer, network layer, supervisory layer, and management layer. Although various layered frameworks have been proposed, they often cater to specific architectural needs. 

\begin{figure}[htbp]
\centerline{\includegraphics[width=3.6 in]{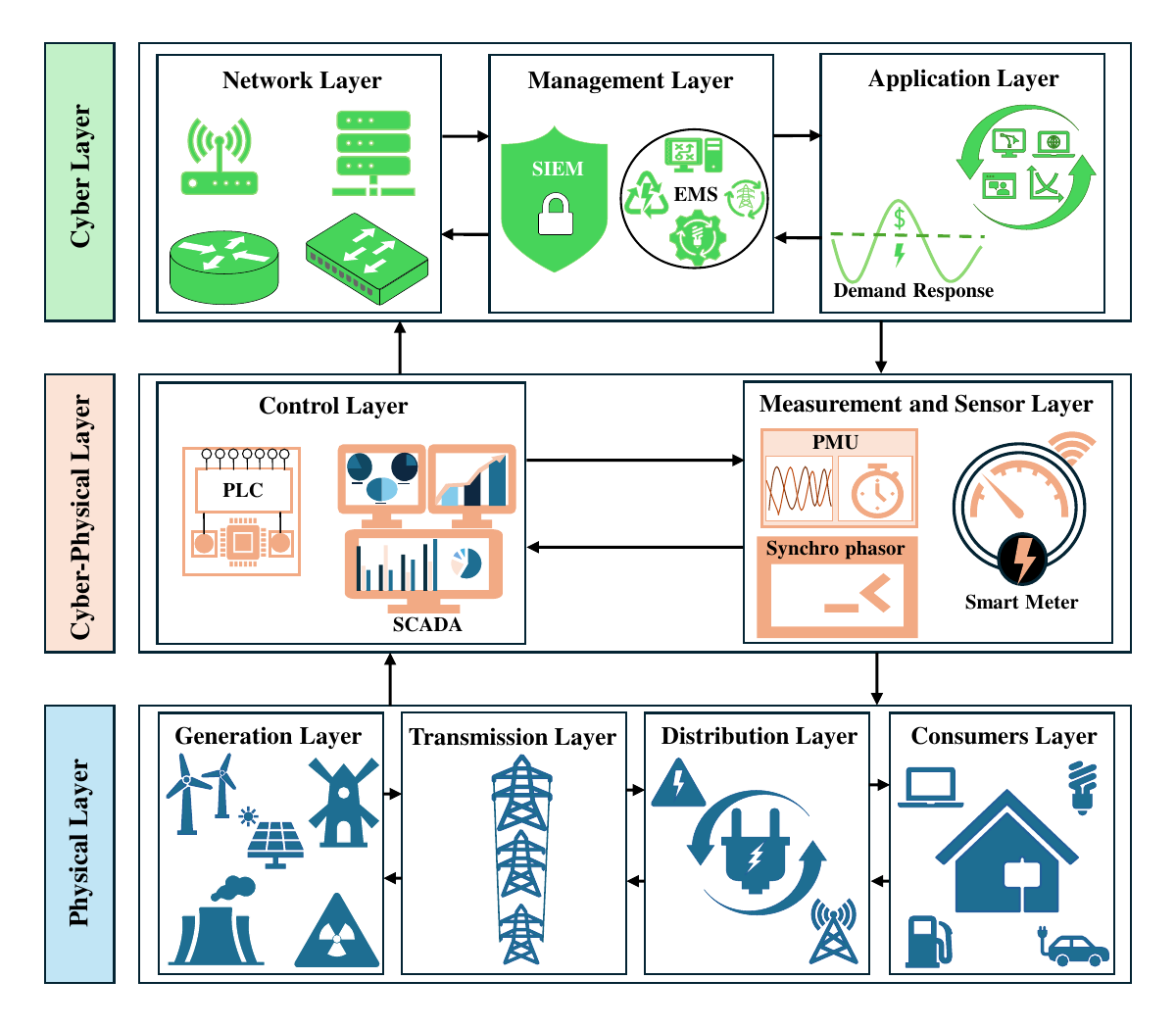}}
\caption{Layers in Smart Grid}
\label{figure:layers}
\end{figure}

In this study, we introduce a more generalized framework with three primary layers, each with its own specific sub-layers. Figure \ref{figure:layers} shows the three primary layers of the proposed framework.

\subsubsection{Physical Layer}

We define the physical layer as the foundation of the smart grid, covering the generation, transmission, distribution, and consumption of electrical power. The physical layer ensures that electricity is generated and transported efficiently and reliably across the grid to meet the demands of diverse consumers. In our definition, the physical layer includes the following sub-layers:

\begin{itemize}
			
\item {\it Generation Layer} involves the production of electrical energy from various sources, such as fossil fuels, nuclear power, and renewable energy sources like solar, wind, and hydroelectric power.

\item {\it Transmission Layer} is responsible for the high-voltage transportation of electrical power from the generation sources to distribution substations.

\item {\it Distribution Layer} manages the lower voltage distribution of electricity from substations to consumers, ensuring that electricity reaches the end-users safely and reliably.

\item {\it Consumption Layer} focuses on the end-users of electricity, including households, industries, and commercial establishments. It encompasses the devices and appliances that utilize electrical energy.

\end{itemize}

\subsubsection{Cyber-Physical Layer}

We define the cyber-physical layer as the intermediary between the physical and cyber layers, connecting the physical components of the smart grid with their digital counterparts. This layer is essential for monitoring, control, and optimization of the smart grid performance. In our definition, the sub-layers within the cyber-physical layer include:

\begin{itemize}

\item {\it Control Layer} facilitates real-time control of the smart grid components. It employs control algorithms in automation devices and deploys communication protocols directed through routers, switches, and gateways. This layer ensures that the grid functions optimally and adapts to changes in demand, effectively regulating the performance of the smart grid.

\item {\it Measurement and Sensor Layer} is integral to the collection and processing of data from various sensors and meters installed across the grid. These sensors provide crucial information on the performance and status of the grid components, enabling the control layer to make informed decisions and adjustments.

\end{itemize}

\subsubsection{Cyber Layer} 

We define the cyber layer as the digital backbone of the smart grid, responsible for the communication, data processing, and management of the grid operations. This layer facilitates the seamless exchange of information and control signals between various grid components, enabling efficient and coordinated operations. In our definition, the sub-layers within the cyber layer include:

\begin{itemize}

\item{\it Network Layer} forms the essential communication infrastructure, ensuring seamless data transmission and exchange among various components of the smart grid. This layer is integral to data flow management and plays a key role in maintaining data integrity and security. It harnesses a variety of modern wireless and wired communication technologies, integrating them to establish a robust and resilient network that forms the backbone of the smart grid communication system. 
			
\item {\it Management Layer} focuses on the supervision and management of smart grid operations to ensure effective and efficient functioning. The management layer is integral for data processing, monitoring, diagnostics, and decision-making processes to ensure the secure, efficient, and reliable operation of smart grids. 

\item {\it Application Layer} is concerned with the specific applications and services built on top of the smart grid infrastructure. It involves various software applications and user interfaces that are designed to enhance the overall efficiency, reliability, and sustainability of the energy sector.

\end{itemize}

\subsection{Components in Smart Grid}		

In smart grids, various components work together to ensure efficient functioning and management of modern power systems. These components can be broadly categorized into three classes: 1) communication, data management and control components; 2) energy generation, storage and efficiency components; 3) security components.

\subsubsection{Communication, Data Management and Control Components}

These components form the communication backbone and control mechanisms of the smart grid. 

\begin{itemize}

\item {\it Advanced Metering Infrastructure (AMI)} is a pivotal component of smart grids, enabling two-way communication between utilities and customers \cite{strategy2008advanced}. AMI facilitates the collection and analysis of energy usage data, supporting demand response and energy management. 

\item {\it Distribution Management Systems (DMS)} streamline the management of distribution networks, enhancing efficiency and reliability \cite{cassel1993distribution}. DMS integrates various functions such as fault location, isolation, and service restoration, improving overall operational performance.

\item {\it Distribution Automation (DA) systems} optimize the operation of distribution networks using advanced sensors, communication technologies, and control devices \cite{shirmohammadi1996distribution}. DA systems enable real-time monitoring and control of the distribution network. 

\item {\it Phasor Measurement Units (PMUs)} provide real-time monitoring of power systems, measuring voltage, current, and frequency at high resolutions \cite{wilson1994pmus}. PMUs enhance the ability to detect and respond to grid disturbances, improving system stability and reliability. 

\item {\it Remote Terminal Units (RTUs)} collect data from field sensors, convert it to digital form, and transmit it to the central system \cite{tavora1979remote}. They also receive and execute control commands, enabling remote monitoring and control of grid components.

\item {\it Supervisory Control and Data Acquisition (SCADA) systems} monitor and control critical infrastructure in the power grid, providing real-time data on system states \cite{gaushell1987supervisory}. SCADA systems enable operators to make informed decisions and take timely actions to maintain grid stability and security.

\item {\it Wide Area Monitoring Systems (WAMS)} monitor power systems over large geographical areas, offering real-time data on stability and performance \cite{mittelstadt1996doe}. WAMS helps in the early detection of potential grid issues, allowing for proactive measures to prevent large-scale outages.

\end{itemize}

\subsubsection{Energy Generation, Storage, and Efficiency Components}

These components are integral to energy production, storage, and management of energy demand. 

\begin{itemize}

\item {\it Microgrids} are small-scale power grids that can operate independently or in conjunction with the main grid, improving resilience and energy efficiency \cite{lasseter2004microgrid}. 

\item {\it Distributed Energy Resources (DERs)} include renewable energy sources like solar panels and wind turbines, which contribute to a sustainable and diverse energy mix \cite{akorede2010distributed}. 

\item {\it Energy Storage Systems (ESS)} help balance supply and demand by storing excess energy and releasing it during peak demand periods \cite{ribeiro2001energy}. 

\item {\it Demand Response (DR) programs} allow utilities to adjust energy consumption during peak demand periods to alleviate grid stress \cite{zhang2012demand}.

\end{itemize}

\subsubsection{Security Components}

These components establish critical defenses to secure the grid from cyber threats and unauthorized access. 

\begin{itemize}

\item {\it Intrusion Detection Systems (IDS)} actively monitor and identify suspicious activities, protecting the grid from potential cyber attacks. 

\item {\it Firewalls} establish a barrier between the trusted internal network of the smart grid and potential threats from outside networks. 

\end{itemize}
		
While the integration of new components such as AMI, PMU, and SCADA indeed enhances the efficiency and reliability of smart grids, it also brings new security challenges. This is because these components rely on ICT infrastructure, making them vulnerable to potential cyber attacks.

\subsection{Communication Protocols in Samrt Grids}

In smart grids, communication protocols are essential to facilitate efficient data exchange and ensure seamless interaction between various components and devices. DNP3, Modbus, and MQTT are the most prominent communication protocols adopted in smart grids. 

\begin{itemize}

\item {\it Distributed Network Protocol 3 (DNP3)} is widely used in SCADA systems \cite{majdalawieh2006dnpsec}. It supports features such as authentication, encryption, and data integrity to ensure the confidentiality, integrity, and availability of sensitive information and control commands. 
		
\item {\it Modbus} is a common communication protocol in industrial control systems and smart grid operations, used for monitoring and controlling devices like intelligent electronic devices (IEDs), programmable logic controllers (PLCs) and RTUs \cite{swales1999open}. Due to its age and simplicity, Modbus lacks robust security features such as authentication, encryption, and data integrity mechanisms, leaving the smart grid vulnerable to cyber attacks.
		
\item {\it Message Queuing Telemetry Transport (MQTT)} is a lightweight, publish-subscribe messaging protocol designed for low-bandwidth and high-latency networks \cite{standard2014mqtt}. MQTT offers inherent security features such as transport layer security (TLS) and secure sockets layer (SSL) encryption, which provide secure channels for data transmission. However, its lightweight nature may expose it to potential vulnerabilities such as weak authentication, insecure default settings, and susceptibility to denial of service (DoS) attacks. 	

\end{itemize}

In the subsequent sections, we will explore various network attacks targeting different layers of smart grids and examine emerging mitigation strategies to defend against these threats.

\section{Emerging Attacks in Smart Grid}
\label{section:attack}

In this section, we focus on understanding various types of attacks that undermine the security of smart grids. While attacks on power grids have been extensively investigated, our interest is primarily in the attacks associated with the newly incorporated smart grid components. Specifically, we categorize these attacks into three main groups: {\it cyber layer attacks}, which aim at the cyber infrastructure of smart grids to disrupt operations and compromise data privacy; {\it cyber-physical layer attacks}, which target the control and sensor systems in smart grids, potentially causing operational failures and physical consequences; {\it coordinated attacks}, which combine cyber and physical tactics to simultaneously attack various grid components, undermining the core security principles of confidentiality, integrity, and availability. 
		
For each category, we will discuss different attacks from three security perspectives: 

\begin{itemize}

\item {\it Confidentiality}, which involves safeguarding against unauthorized access, use, or disclosure of sensitive information.

\item {\it Integrity}, which ensures the accuracy, completeness, and uncorrupted state of data and system components.

\item {\it Availability}, which pertains to the system capability to provide timely access to resources and services when needed. 

\end{itemize}

\subsection{Cyber Layer Attacks}

Cyber layer attacks in smart grids can be defined as unauthorized or malicious activities targeting various components within the cyber layer of smart grids (including network, management, and application sub-layers). These attacks utilize various tactics like malware, phishing, ransomware, and DoS attacks, each with distinct objectives such as breaching sensitive data, exploiting financial gain, or disrupting network functioning. Table \ref{table:cyber} categorizes cyber layer attacks based on malicious objectives that impact the security aspects of confidentiality, integrity, or availability.

		\begin{table}[h]
  \renewcommand\arraystretch{1.5}
			\caption{Taxonomy of Attacks in the Cyber Layer}
			\label{table:cyber}
			\begin{tabular}{|m{2cm}<{\centering}|p{3.5cm}|m{2cm}<{\centering}|}
				\hline
				\centering{\textbf{Security Impact}} & \centering{\textbf{Cyber Attack}} & \textbf{Reference} \\ \hline
				\multirow{5}*{Confidentiality} & Eavesdropping & \cite{baig2013analysis,valli2012eavesdropping,tyav2022comprehensive}\\ \cline{2-3} 
				~& Man-in-the-middle & \cite{gunduz2018analysis,wlazlo2021man,conti2016survey}\\ \cline{2-3} 
				~& Unauthorized Access & \cite{komninos2014survey,yang2011impact,rawat2015cyber,sun2018cyber} \\ \cline{2-3} 
				~& Reconnaissance & \cite{el2018cyber,bristow2008modscan,gonzalez2008passive}\\ \cline{2-3} 
				~& Message Replay & \cite{huitsing2008attack,radoglou2020implementation,aloul2012smart,baig2013analysis}\\ \hline
				\multirow{9}*{Integrity} & Supply Chain Attacks& \cite{duman2019modeling,rao2021iot,businessweek2018big,wolff2021navigating}\\ \cline{2-3} 
				~& Command Manipulation & \cite{musleh2019survey,cherepanov2017industroyer}\\ \cline{2-3} 
				~& Code Manipulation & \cite{musleh2019survey,bencsath2012cousins}\\ \cline{2-3} 
				~& Malware Injection & \cite{khan2016threat}\\ \cline{2-3} 
				~& Rogue Node & \cite{alrawais2017fog,zaidi2015host,sahu2017detection}\\ \cline{2-3} 
				~& Broadcast Message Spoofing & \cite{huitsing2008attack}\\ \cline{2-3} 
				~& Topology Attacks & \cite{kim2013topology}, \cite{liu2016local}\\ \cline{2-3} 
				~& Sybil & \cite{levine2006survey,newsome2004sybil,sarigiannidis2015detecting}\\ \cline{2-3} 
				~& Byzantine & \cite{ross2013using, lamport2019byzantine}\\ \hline
				\multirow{2}*{Availability} & DOS/DDOS & \cite{antonakakis2017understanding,huseinovic2020survey, ortega2023review, diaba2023proposed, torres2022icad}\\ \cline{2-3} 
				~& Response Delay & \cite{radoglou2020implementation,huitsing2008attack,moussa2015detection}\\ \hline
			\end{tabular}
		\end{table}

\vspace{0.15cm}		
\subsubsection{Confidentiality}

\underline{Eavesdropping:} The attack (also known as traffic analysis or packet sniffing) refers to the interception and analysis of network traffic to gain sensitive information \cite{baig2013analysis}. In smart grids, eavesdropping may expose critical data to attackers including energy consumption patterns, pricing information, control messages, etc. Encryption and secure communication protocols can help mitigate the risk of eavesdropping in smart grids \cite{valli2012eavesdropping,tyav2022comprehensive}. 
			
\underline{Man-in-the-middle (MITM):} These attacks occur when an adversary intercepts communication between two parties, potentially intercepting, modifying, or injecting packets without the original participants' knowledge \cite{gunduz2018analysis}. In smart grids, a successful MITM attack could lead to unauthorized access to sensitive data, FDI , or the manipulation of control messages \cite{wlazlo2021man}. Defense strategies against MITM attacks include cryptographic-based prevention (e.g., TLS), anomaly mitigation, and anomaly detection \cite{conti2016survey}.
		
\underline{Unauthorized Access:} Such attack aims to grant illegitimate access to critical systems (e.g., control centers, energy management platforms), resulting in the tampering of energy usage data, disruption of grid operations, or breach of customer information in smart grids \cite{komninos2014survey}. One prevalent form of unauthorized access is password pilfering, which steals users' passwords to gain unauthorized access to sensitive information or systems~\cite{yang2011impact}. Preventing unauthorized access in smart grids requires a combination of strong authentication, access control policies, firewalls, and regular security audits \cite{rawat2015cyber, sun2018cyber}.
		
\underline{Reconnaissance Attacks:} These attacks can either actively scan the network to identify critical infrastructures or passively monitor the network traffic to identify potential vulnerabilities of grid's operations, which serve as preliminary steps for more advanced cyber aggression~\cite{gonzalez2008passive}. A specific instance of reconnaissance, known as Modbus Network Scanning, scans the network to identify devices operating with the Modbus protocol, and further launches attacks on the target devices \cite{bristow2008modscan}. 

\underline{Message Replay:} Such attack, also known as a replay attack, is a type of cyber-attack where an attacker intercepts and retransmits a valid data transmission, often with malicious intent. One such example in smart grids is the baseline response replay attack, where the attacker captures legitimate responses from a Modbus device, and then replays these responses at a later time to either mislead the system or conceal malicious activities \cite{huitsing2008attack, radoglou2020implementation, aloul2012smart}. Such attacks can lead to incorrect system states being reported, potentially leading to misguided operational decisions or hidden malicious activities. Replay attacks can be mitigated by using various security measures such as time-stamping, sequence numbering, or cryptographic techniques like nonce or digital signatures \cite{baig2013analysis}.
		
\subsubsection{Integrity}

\underline{Supply Chain Attacks:} These attacks aim to compromise a smart grid by targeting various stages of a product life cycle (e.g., manufacturing, distribution, and maintenance)\cite{duman2019modeling}. Particularly, hardware supply chain attacks aim to tamper with or manipulate hardware components during the production or distribution, potentially introducing backdoors or vulnerabilities into smart grid equipments \cite{rao2021iot}. Also, in 2018, Bloomberg reported that malicious microchips were inserted during the manufacturing process, compromising many companies \cite{businessweek2018big}. On the software side, supply chain attacks focus on compromising the development, distribution, or update processes of software, potentially leading to the installation of malicious code within smart grids. A notable example is the SolarWinds attack in 2020, where attackers embedded a malicious script into the software updates, resulting in breaches of multiple U.S. government agencies and companies~\cite{wolff2021navigating}.

\underline{Command Manipulation:} Such attack aims to alter the control commands in smart grids, potentially leading to inappropriate operations, equipment damage, or service disruptions\cite{musleh2019survey}. A real-world example is Industroyer, a highly sophisticated piece of malware, which was specifically designed to target Ukraine's power grid. In 2016, it successfully manipulated control commands, resulting in a significant power outage \cite{cherepanov2017industroyer}.
			
\underline{Code Manipulation:} This attack targets the software running on smart grid systems, altering its behavior to introduce vulnerabilities or facilitate unauthorized access \cite{musleh2019survey}. A notable example of code manipulation is the malware Flame (sKyWIper), which was discovered in 2012. It primarily targeted computer systems in the Middle East, and was intricately designed to manipulate code in order to steal sensitive information and monitor user activities~\cite{bencsath2012cousins}. 
			
\underline{Malware Injection:} The attack aims to insert malicious software into systems, potentially disrupting their functionality, stealing sensitive information, or facilitating further attacks. A real-world instance of such an attack is the malware known as BlackEnergy, which evolved from a simple DDoS platform to a highly sophisticated, plug-in based malware capable of targeting critical infrastructures. By exploiting vulnerabilities in widely-used communication standards, BlackEnergy allows attackers to gain unauthorized access, steal sensitive information, and disrupt essential operations, posing a significant risks to critical systems and infrastructure \cite{khan2016threat}.  

\underline{Rogue Node:} Such attack aims to intercept, modify, or inject false data, commands, or messages, directly causing security breaches and system instability of smart grids \cite{alrawais2017fog}. Such attacks can be mitigated by incorporating robust authentication mechanisms that prevent unauthorized users from accessing system assets \cite{zaidi2015host, sahu2017detection}.		

\underline{Broadcast Message Spoofing:} The attack specifically targets smart grids using the Modbus protocol, where an attacker forges a broadcast message that appears to originate from a legitimate source, potentially distributing incorrect commands across the grid \cite{huitsing2008attack}. The objective of the attack is to deceive the system into acting on fraudulent commands, emphasizing the critical importance of secure and verifiable communication within the smart grid infrastructure.
			
\underline{Topology Attacks:} These attacks target the structure and configuration of a network to disrupt the information flow between its components and nodes \cite{kim2013topology,liu2016local}. Such attacks provide false or misleading data of the grid topology (e.g., fake routing tables or configuration files) to control systems, potentially disrupting the communication in smart grids. 
			
\underline{Sybil:} The attack occurs when a malicious node in a network illegitimately claims multiple identities \cite{levine2006survey,newsome2004sybil}. By controlling a large number of identities, a Sybil attack can disproportionately impact the network operations, manipulate data aggregation, disrupt communication, or undermine trust-based schemes \cite{sarigiannidis2015detecting}. Such attack is particularly threatening to distributed systems (e.g., wireless sensor networks in smart grids), where nodes often rely on information from other nodes to make decisions or perform tasks. 
			
\underline{Byzantine:} This attack occurs in distributed systems when one or more nodes behave maliciously or unpredictably, undermining the system reliability \cite{ross2013using}. As smart grids rely on distributed control systems, malicious nodes can disseminate conflicting information to various parts of the system, leading to inaccurate data aggregation, communication disruption, or incorrect system decisions \cite{lamport2019byzantine}.

\subsubsection{Availability}

\underline{DoS/DDoS Attacks:} These attacks have different operational tactics but share a common objective - to overwhelm a target system or network, rendering it unavailable or inaccessible. A DoS attack leverages one device to flood a target with excessive traffic or requests, whereas a DDoS attack coordinates multiple devices (e.g., botnets) to generate the deluge of requests. In smart grids, such attacks may target key components such as control systems or communication networks, potentially leading to significant disruptions in power distribution and management \cite{huseinovic2020survey}. Mirai botnet is a real-world example of botnet-driven DDoS attacks, where the attacker exploits vulnerable IoT devices to launch large-scale attacks against various internet services \cite{antonakakis2017understanding}. To mitigate these attacks, it is essential to implement robust network security protocols, such as deploying IDSs, or adopting traffic filtering techniques.

\underline{Response Delay Attack:} Such attack directly affects service availability by intentionally increasing the latency in the response of a Modbus device \cite{radoglou2020implementation}. The resultant delays can disrupt the real-time operation of the system, leading to synchronization issues, incorrect system states, and even unavailability of control responses. For example, if the command to boot power production in response to demand surge is significantly delayed, it could result in power shortages and outages.

\textbf{Research challenges against cyber-layer attacks:} Though cryptography based authentication, encryption and IDSs can help mitigate many cyber-layer attacks, defending against them presents various challenges due to the complexity and critical nature of the smart grid infrastructure. The diverse range of attack vectors requires the defense strategies to be both comprehensive and highly flexible. As smart grids are interconnected, a cyber-layer attack may not only compromise the integrity of the network, management, and application sub-layers, but also cascade to other layers, resulting in more extensive damage. It is necessary to deploy a suite of security strategies that can work cohesively to provide robust protection across all layers of the smart grids. Also, the continually evolving attacks necessitate ongoing updates and adaptation of security protocols and systems. Further, implementing effective security strategies should consider the potential false positives and the imperative for immediate and rapid response to threats, all while maintaining the grid's operational integrity. These challenges underscore the importance of developing dynamic, resilient, and intelligent security solutions that can mitigate the ever-changing threats to the cyber layer of smart grids.

\subsection{Cyber-physical Layer Attacks}

Cyber-physical layer attacks exploit vulnerabilities in the cyber-physical layers (including control, measurement and sensor sub-layers), directly or indirectly impacting the smart grid's functionality. For example, an attack could disrupt control algorithms leading to equipment dysfunction or electricity misrouting. Alternatively, an attacker could compromise the integrity of sensor data, undermining the grid reliability and resulting in power outages. Table \ref{table:cyberphy} outlines the classification of cyber-physical attacks based on their malicious objectives. 
\begin{table}[h]
\renewcommand\arraystretch{1.5}
	\caption{Taxonomy of Attacks in the Cyber-physical Layer}
	\label{table:cyberphy}
	\begin{tabular}{|m{2cm}<{\centering}|p{4cm}|m{1.5cm}<{\centering}|}
		\hline
		\centering{\textbf{Security Impact}} & \centering{\textbf{Cyber-Physical Attack}} & \textbf{Reference} \\ \hline
		\multirow{2}*{Confidentiality} & Side Channel Attack & \cite{standaert2010introduction,le2008overview,gandolfi2001electromagnetic,jiang2017novel,wang2020mitigating,cilio2013mitigating}\\ \cline{2-3} 
		~& Insider Attacks &  \cite{bao2015blithe}\\ \hline
		\multirow{12}*{Integrity} & GPS Spoofing and Time Synchronization Attacks &  \cite{fan2017synchrophasor,zhang2013time,jafarnia2012gps,gong2012gps}\\ \cline{2-3} 
		~& False Data Injection (FDI) & \cite{liang2016review,liu2011false,bobba2010detecting,xie2010false,teixeira2010cyber,zhang2018distributed,yu2015blind}\\ \cline{2-3}
		~& Attacks on Automatic Generation Control (AGC) &  \cite{sridhar2014model,ameli2018attack}\\ \cline{2-3}
		~& Attacks on Voltage Control &  \cite{isozaki2015detection,chen2022distributed,bhusal2021detection,teixeira2015voltage,ju2018adversarial, langer2016analysing}\\ \cline{2-3}
		~& Demand Side Management Attacks & \cite{giraldo2016integrity, tan2013impact}\\ \cline{2-3}
		~& Switching Attacks & \cite{liu2013framework,liu2011switched,liu2011class,liu2012smart,liu2012coordinated,wu2019optimal}\\ \cline{2-3}
		~& Load Redistribution & \cite{lee2019vulnerability,xiang2016power,yuan2012quantitative,yuan2011modeling}\\ \cline{2-3}
		~& Attacks in Vehicle-to-grid (V2G) & \cite{saxena2017network,zhang2013securing,han2016privacy}\\ \cline{2-3}
		~& Aurora & \cite{zeller2011myth,srivastava2013modeling,zeller2011common,he2016cyber}\\ \cline{2-3}
		~& Advanced Persistent Threats (APTs) & \cite{shakarian2011stuxnet,farwell2011stuxnet,zetter2015countdown,chen2010stuxnet,falliere2011w32,hemsley2018history,response2014dragonfly,wangen2015role,hentunen2014havex,lee2017crashoverride,slowik2019crashoverride,bindra2017securing,greenberg2017crash}  \\ \cline{2-3}
		~& Source ID Mix Attacks & \cite{zhang2021smart, cui2019spatio, liu2020model}\\ \cline{2-3}
		~& Rogue Interloper & \cite{east2009taxonomy,siddavatam2015security}\\ \hline
		Availability & Puppet Attacks & \cite{yi2016puppet,jacoba2023cybersecurity}\\ \hline
	\end{tabular}
\end{table}	
		
\subsubsection{Confidentiality}

\underline{Side Channel attacks:} These attacks exploit inadvertent information leaks from physical systems to compromise sensitive data. Such attacks usually target devices using cryptographic techniques (e.g., smart meters, PMUs) \cite{standaert2010introduction}. Power analysis attacks, for example, infer secret keys by analyzing the power consumption patterns of a device during cryptographic operations \cite{le2008overview}. Electromagnetic analysis attacks extract sensitive information, such as cryptographic keys, from electromagnetic emissions of a device \cite{gandolfi2001electromagnetic}. Timing attacks exploit variations in operation times to gain sensitive data~\cite{jiang2017novel}. Mitigation strategies such as the adoption of secure hardware designs, and randomization of timing or power consumption patterns can be deployed to defend against such attacks. \cite{wang2020mitigating, cilio2013mitigating}. 

\underline{Insider Attacks:} Such attacks are perpetrated by insiders who have authorized access to the network and systems within the smart grid. An insider could be an employee, contractor, or even a business partner for their personal gains or malicious intents. They have the sufficient knowledge of the system to bypass the security enforcement, potentially resulting in data theft, system damage, or operational disruption. IDS can be deployed to identify suspicious activities of insider attacks \cite{bao2015blithe}.

\subsubsection{Integrity}

\underline{Switching Manipulation Attacks:} In switching manipulation attacks, an adversary aims to manipulate the state of switches (e.g., circuit breakers, transfer switches) to disrupt normal grid operation and cause physical damage~\cite{liu2013framework,liu2011switched,liu2011class,liu2012smart, liu2012coordinated}. By modifying the state of switches, the attacker can alter power flows, cause overloads, or even isolate parts of the grid, leading to significant operational challenges and risks \cite{wu2019optimal}.

\underline{Load Redistribution Attacks:} In such attacks, an adversary aims to manipulate the control system to alter the distribution of electrical loads across the network \cite{lee2019vulnerability,xiang2016power}. The attacks can be accomplished by falsifying data or dispatching malicious commands to control systems \cite{yuan2012quantitative}. By redistributing the load inappropriately, the attacker can cause physical equipment to overload and potentially fail, leading to blackouts or other disruptions of service \cite{yuan2011modeling}.
			
\underline{Attacks in Vehicle-to-grid (V2G):} V2G technologies allow two-way energy exchange between electric vehicles (EVs) and the power grid, enabling EVs to act as distributed energy resources \cite{saxena2017network}. In the attack scenario, a malicious actor might hack into the control systems of autonomous vehicles or the V2G communication interface to manipulate power flow, spoof charging/discharging commands, or disrupt grid balance by injecting false data \cite{zhang2013securing}. Such attacks could lead to local and wider grid instabilities, causing power quality issues or even blackouts. Moreover, the vehicle's functionality could be compromised, posing safety risks \cite{han2016privacy}.

\underline{Aurora:} The attack targets the electrical generators of a power grid and aims to de-synchronize the phase of a generator from the power grid (e.g., opening and closing circuit breakers to alter the physical process of the power system). It dynamically accelerates or decelerates the generator, imposing mechanical stress and eventually damaging or destroying the generator. Aurora attack is first discovered by the Idaho National Laboratory in the United States in 2007. It is particularly concerning because it can cause physical damage to critical infrastructure, and potentially lead to widespread power outages \cite{zeller2011myth,srivastava2013modeling,zeller2011common,he2016cyber}. 

\underline{Advanced Persistent Threats (APTs):} APTs are a category of prolonged and targeted cyber-attacks designed to gain unauthorized access to a network and remain undetected for an extended period. The primary goal of most APT attacks is to achieve and maintain ongoing access to the targeted network for monitoring activities and breaching sensitive data. STUXNET, uncovered in 2010, is a specific example of APTs capable of causing physical damages to cyber-physical systems. The malicious worm targets the programmable logic controllers (PLCs) within the supervisory control and data acquisition (SCADA) system of Iran nuclear infrastructure \cite{shakarian2011stuxnet}. It exploited vulnerabilities to alter the frequency of the electrical current powering the centrifuges, resulting in physical damage \cite{farwell2011stuxnet}. Havex is another examples of APTs. Developed by the Dragonfly cyber espionage group, Havex is a Remote Access Trojan (RAT) that exploits vulnerabilities in the energy sector, particularly in the United States and Europe. It has been deployed in attacks on Industrial Control Systems (ICS), with a focus on compromising critical infrastructures \cite{hentunen2014havex, response2014dragonfly, wangen2015role}. CrashOverride, also known as Industroyer, is malware employed on attack on the Ukrainian power grid in 2016 \cite{lee2017crashoverride,slowik2019crashoverride}. It is the first malware designed specifically to attack electric grids, which can directly control electricity substation switches and circuit breakers \cite{bindra2017securing}. All these attacks highlight the potentially severe consequences that APTs can have on the interconnected smart grid systems \cite{zetter2015countdown,chen2010stuxnet,falliere2011w32}.
			
\underline{Source ID Mix:} Source ID mix is a type of data spoofing attack that targets wide-area measurement systems in smart grids. Instead of altering the actual measurements, source ID mix tampers with the data from PMUs by confusing their source identification tags. Such attacks can compromise the operational reliability of the network, potentially leading to incorrect actions based on faulty data interpretations. To defend against such attacks, a more robust authentication scheme is required to enforce data authenticity and integrity in power systems \cite{cui2019spatio, liu2020model}. 

\underline{Location and Time Synchronization Attacks:} These attacks are significant threats to smart grid systems, exploiting their reliance on precise timing for crucial operations \cite{fan2017synchrophasor, zhang2013time}. For example, GPS Spoofing transmits counterfeit GPS signals to mislead receivers, potentially causing system instability or incorrect operational decisions, especially in the synchronization of PMUs \cite{jafarnia2012gps}. Similarly, Time Stamp Attacks (TSAs) interfere with precise GPS timing, disrupting functionalities such as transmission line fault detection and event location estimation \cite{gong2012gps}. To protect against these timing-dependent threats, multi-antenna-based algorithms have been proposed for quick GPS spoofing detection.
   
\underline{False Data Injection (FDI):} These attacks, which rank among the most prevalent attacks in the smart grid, aim to disrupt system operations by injecting false data \cite{liang2016review}. Such attacks can mislead automated systems and operators by introducing erroneous data into PMUs, leading to incorrect operational decisions. State estimators and converters are particularly vulnerable to FDI attacks, which can result in inaccurate grid state estimates, inappropriate power distribution decisions, and potential disruptions or inefficiencies \cite{liu2011false, bobba2010detecting, xie2010false, besati2023new}. FDI attacks can also compromise protection systems designed to detect and isolate grid faults. If such systems are compromised, their ability to accurately detect faults may be impaired, potentially leading to prolonged outages or equipment damage \cite{teixeira2010cyber}. Additionally, during load balancing, attackers may introduce false data about the load on different grid parts, causing flawed power distribution decisions and possibly resulting in overloads or underloads in certain areas \cite{zhang2018distributed}. A variation of the FDI attack is the blind FDI attack, where the adversary injects false data without comprehensive system knowledge. Blind FDI attackers can still cause significant disruption despite their limited system understanding \cite{yu2015blind}. 
			
\underline{Attacks on Automatic Generation Control (AGC):} Attacks on AGC pose significant risks to the stability and reliability of smart grids. The AGC system maintains a real-time balance between generation and load by continually adjusting power output from various generators. If an attacker manages to manipulate the AGC, they can induce frequency instability, overloading or underloading of generators, or even cause blackouts. This could be achieved by injecting false data, tampering with command signals, or exploiting vulnerabilities in the communication network \cite{sridhar2014model, ameli2018attack}.
			
\underline{Attacks on Voltage Control:} In such attacks, adversaries manipulate control systems to alter voltage levels within the grid. They can be achieved by providing false data or sending malicious commands to voltage control devices. Such attacks are particularly concerning because it can lead to physical damage and cascading failures in the power system due to voltage instability \cite{isozaki2015detection}.
			
\underline{Demand Side Management Attacks:} Demand Side Management Attacks, such as Pricing Attacks, aim to manipulate the pricing signals for customers in a demand response system. In pricing attacks, an attacker could falsify price signals to make electricity seem more expensive or cheaper than it actually is. This could lead customers to adjust their energy usage based on false information, leading to instability of grid, financial losses, or even physical damage due to unexpected load changes \cite{giraldo2016integrity,tan2013impact}.
			
\underline{Rogue Interloper Attacks:} A Rogue Interloper attack on smart grids occurs when an unauthorized device or entity infiltrates the network. The rogue device, through identity spoofing or other forms of deception, presents itself as a legitimate component within the smart grid network. Once accepted as part of the system, it can engage in various malicious activities, such as sending false data or commands, disrupting network communication, or even assuming control over certain operations. The potential consequences of such an attack are significant, leading to operational disruptions or even physical damage to equipment. This highlights the necessity of robust device authentication and stringent access control measures within smart grid networks \cite{east2009taxonomy,siddavatam2015security}.

\subsubsection{Availability}

\underline{Puppet Attacks:} In puppet attacks, an attacker gains control of a legitimate node within a network and uses it to carry out malicious activities. The attacker can manipulate the device to send false data, execute unauthorized commands, or disrupt normal operations. Such attacks can indirectly affect the functionality of the smart grid, for instance, by exhausting the bandwidth needed for the control algorithms to function correctly, leading to equipment malfunction or electricity misrouting. Additionally, they can compromise the network operation, potentially resulting in power outages or damage to physical infrastructure \cite{yi2016puppet}.

\textbf{Research challenges against cyber-physical layer attacks:} Defending against cyber-physical layer attacks in smart grids presents unique challenges due to their direct impacts on physical infrastructure. Unlike cyber-layer attacks, which primarily target network and application, cyber-physical attacks can cause actual physical damage to grid components. These attacks exploit the control and sensor layers, leading to malfunctions in equipment and mismanagement of electricity flow. Maintaining integrity in grids is particularly challenging because many attacks in this layer specifically aim to compromise this security aspect. The primary goal against such attacks is to maintain grid stability and prevent physical damage, even in the face of sophisticated attacks that aim to disrupt the balance between supply and demand in smart grids. The defense requires not only securing the network from unauthorized access but also ensuring the integrity of control commands and the authenticity of sensor data. This necessitates implementing secure communication protocols, real-time anomaly detection systems, and resilient control strategies that can adapt to and compensate for malicious activities. 

\subsection{Coordinated (hybrid) Attacks}

Coordinated attacks utilize a mix of cyber and cyber-physical attack methods to conduct simultaneous or sequential strikes on multiple components within the smart grid \cite{tian2020coordinated}. These attacks can involve complex strategies, such as launching a cyber attack to disrupt communication systems followed by a physical attack on critical infrastructure, or vice versa. Despite their complexity, coordinated attacks are highly effective and difficult to detect because they leverage covert masking tactics to compromise multiple security objectives simultaneously. These tactics can include blending malicious actions with legitimate operations to avoid detection and exploiting timing and sequencing to maximize impact. Coordinated attacks typically compromise multiple aspects of security. Table \ref{table:coordinated_attacks} outlines various types of coordinated attacks, classifying them based on directly compromised security aspects.

\begin{table}[h]
\renewcommand\arraystretch{1.5}
\caption{Taxonomy of Coordinated Attacks in Smart Grid}
\label{table:coordinated_attacks}
\centering
\setlength{\tabcolsep}{4pt}
{\footnotesize
\begin{tabular}{|m{3cm}<{\centering}|m{3cm}<{\centering}|m{1.6cm}<{\centering}|}
\hline
\textbf{Security Impact} & \textbf{Coordinated Attacks} & \textbf{Reference} \\ 
\hline
Integrity & Coordinated Data Injection Attacks & \cite{cui2012coordinated}\\
\hline
Availability & Coordinated Load Changing Attacks & \cite{arnaboldi2020modelling, dabrowski2017grid} \\
\hline
Confidentiality, Integrity & Coordinated Replay Attacks & \cite{deng2017ccpa} \\
\hline
\multirow{3}*{Integrity, Availability} & Coordinated Topology Attacks & \cite{yan2016q, zhang2016physical, li2015bilevel, wang2020coordinated}\\
\cline{2-3}
~ & Line Outage Masking Attacks & \cite{li2016analyzing, case2016analysis, chung2018local, li2015bilevel, liu2016masking, tian2019multilevel} \\
\cline{2-3}
~ & Coordinated DoS Attacks & \cite{tian2020coordinated} \\
\hline
{\shortstack{Confidentiality, Integrity, \\Availability}} & Line Outage Masking Attacks & \cite{soltan2018react, soltan2018expose} \\
\hline
\end{tabular}
}
\end{table}


\underline{Coordinated Topology Attacks:} These attacks compromise smart grid operations by manipulating both the physical and cyber components of the system simultaneously. They manipulate state and topology data to mislead control center about line outages and can result in cascading failures within the grid. The attack strategy includes several steps: First, attackers gather detailed topology information and system state data. Next, they physically trip a transmission line to create a real outage and use false data injection to hide the outage of the tripped line. Finally, the attackers generate a fake outage signal for another line, misleading the control center into rerouting power improperly, which can potentially overload critical lines. By carefully coordinating these actions, attackers make it difficult for operators to distinguish between real and fake outages, leading to incorrect operational decisions and increased grid vulnerability \cite{yan2016q}.


\underline{Coordinated Data Injection Attacks:} The primary goal of these attacks is to mislead the energy management system by injecting data that resembles normal grid operations, resulting in incorrect control decisions and potential large-scale power outages. These cooperative attackers exploit their knowledge of the power network topology and collaborate to inject false data that appears normal to avoid traditional detection mechanisms in smart girds. Quickest Detection has been proposed to promptly detect and mitigate such coordinated false data injections before they can cause significant damage to the power grid infrastructure \cite{cui2012coordinated}.

\underline{Coordinated Load Changing Attacks:} Such attacks aim to coordinate botnets (i.e., a large number of infected IoT devices) to create sudden spikes or drops in energy usage, disrupting the balance between power supply and demand in a smart grid \cite{arnaboldi2020modelling}. Typically, the attackers can manipulate the infected IoT devices to simultaneously turn high-wattage appliances on or off, creating the spikes or drops. The rapid changes can destabilize the power grid, leading to automatic load shedding and potentially causing widespread power outages \cite{dabrowski2017grid}.


\underline{Coordinated Replay Attacks:} These attacks aim to mask the effects of physical attacks on the power grid by replaying recorded normal meter measurements to control centers. To be effective, coordinated replay attacks generally require compromising a large number of branch meters, ensuring the false data appears consistent and avoids detection. An optimized approach to such attacks can evade bad data detection systems by strategically altering only four meter measurements, making them more stealthy and efficient \cite{deng2017ccpa}. Recent  strategies to defend against such attacks focuses on developing multilevel programming models to enhance the interaction representation between the control center and adversaries, thereby improving the detection of overloaded transmission lines \cite{tian2019multilevel}.

			
\underline{Line Outage Masking attacks:} The objective of these attacks is to mislead the control center into believing that the grid is operating normally, despite one or more transmission lines being out of service \cite{li2016analyzing,chung2018local}. These attacks pose a significant threat to the stability and security of smart grids, leading to severe operational risks and widespread power outages\cite{case2016analysis}. To hide the line disruptions, attackers can exploit RTUs and PMUs through malicious circuit breaker manipulation and DDoS attacks, erasing the master boot record and overwhelming call centers \cite{li2015bilevel, liu2016masking, tian2019multilevel}. Continuous monitoring and quick detection are critical to mitigate the risks posed by Line Outage Masking attacks and to maintain the integrity and reliability of the power grid \cite{soltan2018react, soltan2018expose, soltan2019line}. 

			
\underline{Coordinated DoS Attacks:} These attacks target both physical infrastructure and SCADA systems, resulting in severe consequences. The strategy generally begins with an initial physical disruption, such as tampering with power flow meters, followed by DoS attacks to disrupt communication channels between SCADA systems and their sensors or control devices. Such coordination masks the disruption by making critical measurement data unavailable, preventing system operators from detecting the initial damage. Consequently, it leads to incorrect operational decisions and a cascade of failures, significantly impeding the grid's ability to respond effectively \cite{tian2020coordinated}.




\textbf{Research challenges against coordinated attacks:}
It is crucial to develop sophisticated detection mechanisms to defend against coordinated attacks in smart grids. One strategy is the use of Collaborative Intrusion Detection Systems (CIDS), which integrate IDS across different nodes to provide comprehensive monitoring and a unified response strategy \cite{zhou2010survey}. Similarly, the Coordinated Attack Response and Detection System (CARDS) employs a signature-based model for efficient data collection, analysis, and multi-source information correlation \cite{yang2000cards}. Additionally, \cite{cui2012coordinated} improves detection solutions by understanding attacker-defender dynamics, designing distributed detection algorithms, and evaluating the trade-offs between detection speed and reliability. Coordinated attacks can also be identified by analyzing cross-domain attack information \cite{sen2021towards}. \cite{lai2019tri} proposes a tri-level optimization model to defend against coordinated attacks by strategically distributing defensive resources on both the physical and cyber components of the smart grid.

Despite advanced defense mechanisms, addressing coordinated attacks on smart grids remains a significant challenge. First, the complex nature of these attacks, targeting physical, cyber, and cyber-physical layers simultaneously, complicates the development of comprehensive defense strategies. Detecting and mitigating such attacks requires advanced monitoring systems capable of identifying and correlating anomalies across different layers of the grid infrastructure. Second, it is important to ensure scalability of defense strategies for large-scale smart grids, demanding real-time data processing and efficient algorithms. Furthermore, continuous improvement in threat intelligence and information sharing among smart grid operators is essential to defend against emerging coordinated attack strategies. Finally, integrating advanced technologies, such as machine learning, into security frameworks presents both opportunities and challenges. While these technologies can improve detection and response capabilities, they also introduce new vulnerabilities that need to be addressed to ensure robust defense mechanisms.

\begin{table*}[t]
\renewcommand\arraystretch{1.5}
\caption{Taxonomy of game theory detection and mitigation methods}
\centering
\setlength{\tabcolsep}{4pt}
{\footnotesize
\begin{tabular}{|c|c|c|c|c|c|c|}
\hline
 \textbf{Method} & \textbf{Attack}& \textbf{Detection} & \textbf{Mitigation} & \textbf{Strength} & \textbf{Weakness} & \textbf{Reference} \\ \hline
(Non-)cooperative, Non-zero-sum & Intrusions & \ding{51} & \ding{55} & Flexibility, Scalability & Complexity & \cite{alpcan2003game}\\ \hline
Sequential Game Model & Cyber & \ding{55} & \ding{51} & Interdependent Analysis & Simplified Assumptions, Static &\cite{shan2020game}\\ \hline
 Nash Equilibrium & Cyber & \ding{51} &  \ding{55} & Robust & Complexity & \cite{cardenas2012game}\\ \hline
Two-person Zero-Sum & FDI & \ding{51} & \ding{55} & Practical Scenarios& Complexity & \cite{esmalifalak2013bad} \\ \hline
Perfect Bayesian Equilibrium & FDI &  \ding{51} & \ding{55} & Optimization& Not Realistic Assumption & \cite{nikmehr2019game} \\ \hline
 Zero-sum Stochastic \&\ Q-learning & Cyber & \ding{51} & \ding{55}& Stochastic Modeling& Scalability & \cite{alpcan2006intrusion}\\ \hline
Zero-sum Non-cooperative &  Coordinated CP &\ding{51} & \ding{55} & Moving Target Defense & Convergence Time & \cite{lakshminarayana2021moving}\\ \hline
 Attack-Detection Evolutionary & FDI & \ding{51} & \ding{51} & Adaptability & Reliance on Simulations & \cite{zhang2022preventing}\\ 
\hline
 Non-cooperative, Nash Equilibrium & DDoS &  \ding{51} & \ding{51} & High Accuracy-Dynamic & Complexity & \cite{abou2022federated} \\ \hline
 Bayesian Nash Equilibrium & DDoS &  \ding{51} & \ding{51} & Computational Efficiency & Approximation Errors & \cite{wu2018game}\\ \hline
 Zero-sum Matrix &  APT & \ding{55} & \ding{51}& Optimization & Not Convergent & \cite{rass2017defending} \\ \hline 
 Zero-sum & DoS/DDoS& \ding{55} & \ding{51}& Dynamic & Not Generalized & \cite{de2017game}\\ \hline
 Stochastic Game with Q-Learning & APT & \ding{55} & \ding{51} & Adaptability & Performance Limitations & \cite{chung2016game}\\ \hline
 Nash-Cumulative Prospect Theory & APT & \ding{51} & \ding{55} & Performance & Scalability & \cite{xiao2018attacker}\\ \hline
\end{tabular}
} 
\label{table:game}
\end{table*}

\section{Detection and Mitigation}
\label{section:detection}

In this section, we explore innovative approaches to detect and mitigate different network attacks in smart grids. Specifically, we examine the application of four advanced techniques: game theory, graph theory, blockchain, and machine learning. These techniques have demonstrated effectiveness and prevalence in contemporary research. Each technique offers unique advantages — game theory models the adversarial dynamics, graph-based techniques leverage the networked nature of grids, blockchain provides an immutable audit trail, and machine learning enables adaptable and predictive defenses.


\subsection{Game Theory-based Techniques}

Game theory studies how participants make decisions in strategic situations where the outcome for each depends on the actions of others. It can be used to develop strong security measures in smart grids by modeling the complex interactions between different entities, such as utility providers, customers, and potential attackers, and analyzing the various threats and vulnerabilities involved \cite{saad2012game}. This enables the development of robust security strategies that account for the various players' incentives and potential actions.

Game theory has multiple applications in network intrusion detection for smart grids \cite{roy2010survey}. \cite{alpcan2006intrusion} proposes a 2-player zero-sum stochastic security game to model interactions between malicious attackers and an IDS. The authors in \cite{alpcan2003game} develop a decision-making and control algorithms via a game-theoretic framework. They proposed a distributed IDS model and two game-theoretic schemes: a security warning system for real-time network security overviews and a finite game between the attacker and IDS, with analytical solutions for Nash equilibrium in the security game.

Nash equilibrium is a key concept in game theory, where each participant's strategy is optimal given the strategies of all others. This principle ensures defense strategies are stable and resilient. Nash equilibrium can also introduce mixed strategies, where players randomly choose among multiple actions with certain probabilities, introducing unpredictability that complicates attack strategy of adversaries \cite{fadlullah2011survey}. 

Game theory also offers adaptive defensive approaches against dynamic smart grid threats (e.g., APTs) \cite{park2016game,khalid2023recent}. One typical example is defensive deception, which misleads attackers by manipulating their perceptions and beliefs to protect systems and data \cite{zhu2021survey}. Further, differential game models can be applied to dynamic defense processes and trust frameworks for identifying vulnerabilities. The approach significantly enhances APT attack detection's accuracy, speed, and efficiency.

Additionally, game theory is useful in modeling attacks and defenses within smart grids, addressing challenges like data injection attacks \cite{shan2020game}, energy theft in AMI \cite{jiang2014energy, cardenas2012game}, and cyber threats in the electricity market \cite{esmalifalak2013bad}. Moreover, game theory is important in maintaining global network optimization, especially in FDI attacks on networked microgrids \cite{nikmehr2019game}. Another study presents new game-theoretic approaches for corrupted sensor data detection \cite{vamvoudakis2014detection}. It formulates the problem as a zero-sum game with partial information, where the detector aims to minimize the estimation error and the attacker maximizes it, with the goal to develop efficient solutions for optimal detectors.

Table \ref{table:game} summarizes the game theory methods used, the types of attacks detected or mitigated, and the strengths and weaknesses of each method.

\textbf{Research challenge of game theory applications in smart grid security:}
Ensuring the practical implementation and scalability of game-theoretic solutions in real-world smart grids poses several challenges. One major challenge is accurately modeling the complex interactions between various entities, such as utility providers, customers, and attackers, while considering their diverse objectives and strategies. Additionally, developing realistic threat models that account for the dynamic and evolving nature of cyber threats is difficult. Game theory relies heavily on the assumption of rational behavior, often not reflecting real-world events, resulting in a gap between theory and practice. Another challenge is the computational complexity of finding Nash equilibria and optimal strategies within the resource constraints of smart grid devices \cite{fadlullah2011survey}. Scalability issues also increase as smart grids expand, challenging the application of game theory in large-scale smart grid networks \cite{ma2013scalable}. Moreover, designing robust game-theoretic models that can effectively manage errors and adapt to unpredictable player behaviors is essential for practical and reliable implementation in smart grid systems \cite{saad2012game}.


\begin{table*}[t]
\renewcommand\arraystretch{1.5}
\caption{Taxonomy of graph theory detection and mitigation methods}
\centering
\setlength{\tabcolsep}{5pt}
{\footnotesize
\begin{tabular}{|c|c|c|c|c|c|c|}
\hline
\textbf{Method} & \textbf{Attack}&\textbf{Detection} & \textbf{Mitigation} & \textbf{Strength} & \textbf{Weakness} & \textbf{Reference} \\ \hline
 Graph-based Clustering & DDoS & \ding{51} & \ding{55} & Robustness & Manual Initialization & \cite{mingqiang2012graph}\\ \hline
 Attack Graphs & Multi-step Intrusion &\ding{55} & \ding{51} & Practical & Graph Simplifications & \cite{wang2006minimum}\\ \hline
Graph Signal Processing & FDI &\ding{51} & \ding{55} & Measurement Compatibility & Complexity & \cite{drayer2019detection} \\ \hline
 Graph-based Outlier Detection & FDI &\ding{51} & \ding{55} & Adaptability & Data Dependency & \cite{jorjani2020graph}\\ \hline
 Graph Analytic Metric & Pass-the-Hash &\ding{55} & \ding{51} & Quantitative & Complexity & \cite{johnson2013graph} \\ \hline
 Structural Temporal GNN& Anomalous Edges&\ding{51} & \ding{55} & Temporal \&\ Structural & Potential Overfitting & \cite{cai2021structural}\\ \hline
Graph-based Classification & FDI & \ding{51} & \ding{51} & Observability Maintenance & Complexity & \cite{doostmohammadian2021distributed}\\ \hline
Attack Graphs & Multi-stage & \ding{55} & \ding{51} &Scalability & Complexity & \cite{durkota2019hardening}\\ \hline
 Spatial-Temporal GNN & DDoS & \ding{51} & \ding{51} &  High Accuracy & Not Generalized & \cite{cao2021detecting}\\ \hline 
 Dynamic Reachability Graph & Path-the-Hash & \ding{55} & \ding{51} & Dynamic& Computational Resources & \cite{purvine2016graph}\\ \hline
GNN & Infiltration & \ding{51} & \ding{51} & Scalability & Complexity & \cite{gelenbe2020iot} \\ \hline
\end{tabular}
} 
\label{table:graph}
\end{table*}

\subsection{Graph-based Techniques}

Graph theory is the study of mathematical structures that model pairwise interactions between objects, with vertices (nodes) connected by edges (lines) \cite{diestel2010graph}. Multiple graph-based methods have been developed to defend against network threats in smart grids \cite{staniford1996grids, mingqiang2012graph, ammann2002scalable, jha2002two, phillips1998graph}. Recent studies further extend beyond traditional graph theory, incorporating graph neural networks to enhance attack detection and mitigation strategies. These techniques leverage the principles of graph theory to uncover hidden structures and irregularities within data, facilitating efficient attack detection across various domains such as network security, power systems, and fraud prevention \cite{staniford1996grids, mingqiang2012graph, ammann2002scalable, jha2002two, phillips1998graph,swiler2001computer, noel2003efficient,drayer2019detection, jorjani2020graph, kosut2011malicious, kundur2010towards,pourhabibi2020fraud}. 


Graph-based algorithms have shown their potential for intrusion detection in smart grids.
For example, the GrIDS network transforms network activities into event graphs for suspicious activity detection. Similarly, the Local Deviation Coefficient Graph-Based (LDCGB) algorithm improves data labeling and clustering, enhancing discrimination between normal and anomalous data \cite{mingqiang2012graph}. Graph-based algorithms can also be applied to fraud prevention \cite{pourhabibi2020fraud}, analyzing connectivity patterns in communication networks to detect potential fraudulent activities, including energy theft, billing fraud, and FDI attacks. Additionally, graph-based methods have been explored for network hardening, helping to extract efficient flow paths and identify potential vulnerabilities that attackers might exploit \cite{wang2006minimum}. Case studies also demonstrate the effectiveness of graph-based methods in addressing false data injection in smart grids by utilizing Laplacian matrices and Graph Fourier Transforms (GFT) \cite{drayer2019detection, jorjani2020graph}. Further, graph theory-based methods have been used to test the network vulnerability of lateral movements and privilege escalation attacks, preventing potential APTs \cite{johnson2013graph}. Table \ref{table:graph} summarizes the graph theory methods against different attacks in smart grids. 




{\bf Research challenges of graph-based techniques in smart grid security:} Graph-based techniques in smart grid security face several research challenges. One major challenge is the scalability of these methods. As the size of smart grids increases, the amount of data grows exponentially, making it difficult to design a big data-driven, meaningful, and valid graph for real-time data processing and analysis \cite{wang2013cyber,akoglu2015graph}. Another challenge is the accuracy of anomaly detection. The detection accuracy of graph-based methods may significantly decrease due to noisy and sparse datasets \cite{ranshous2015anomaly, yanmei2024enhanced}. These methods must effectively distinguish between normal operational variations and actual security threats to minimize false positives and false negatives. Additionally, it also poses technical and compatibility issues when incorporating graph-based techniques with existing smart grid infrastructure and communication protocols. Moreover, the dynamic nature of smart grids, with constantly changing topologies and new devices being added, requires adaptive and robust graph-based models that can evolve with the system \cite{cai2021structural, akoglu2015graph}. Finally, it is critical to protect the privacy and security of the data used in graph-based methods, as these techniques rely on extensive data collection and analysis, raising concerns about potential data breaches and misuse.


\subsection{BlockChain}
  
Blockchain technology is a database system that allows for transparent information sharing over a network by organizing data into sequentially linked blocks. Initially designed for Bitcoin transactions \cite{nakamoto2008bitcoin}, blockchain has been actively applied in broad applications, especially toward the increased operation efficiency and security of smart grids \cite{zhuang2020blockchain}. It uses consensus algorithms, immutable ledgers, and decentralization to provide a secure platform for data sharing across sectors \cite{winter2018advantages}. Blockchain technology enhances security by detecting bad data, increasing resilience against network attacks, and preventing data manipulation and tampering \cite{yue2017big, mylrea2017blockchain, kurt2019secure, ghiasi2023comprehensive}.


		\begin{figure}[htbp]
			\centerline{\includegraphics[width=0.5\textwidth]{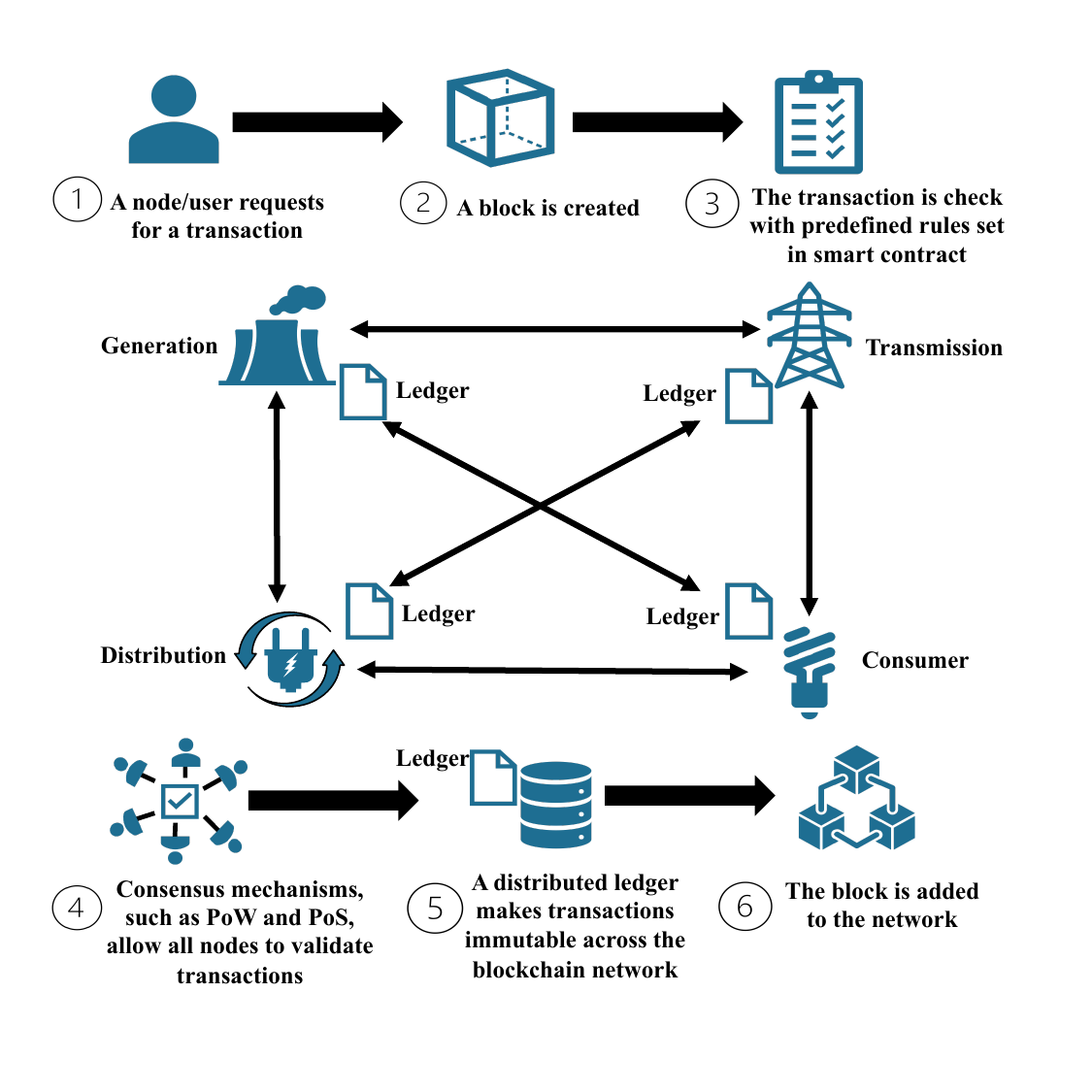}}
			\caption{A workflow illustrating how blockchain technology improves smart grid security.}
			\label{figure:blockchain}
		\end{figure}

\begin{table*}[!t]
\renewcommand\arraystretch{1.5}
\caption{Taxonomy of blockchain theory detection and mitigation methods}
\centering
\setlength{\tabcolsep}{3.9pt}
{\footnotesize
\begin{tabular}{|c|c|c|c|c|c|c|}
\hline
\textbf{Method} & \textbf{Attack}& \textbf{Detection} & \textbf{Mitigation} & \textbf{Strengths} & \textbf{Weakness} & \textbf{Reference}\\ \hline
PoW Consensus &DDoS& \ding{55} & \ding{51} & Decentralization & Complexity & \cite{yue2017big}\\ \hline
 Keyless Signature Infrastructure  &Cyber& \ding{51} & \ding{51} & Real-Time Transactions & Interoperability, Latency & \cite{mylrea2017blockchain}\\ \hline
\parbox[c][2.3em]{4.3cm}{\centering Blockchain-based Secure Distributed Dynamic State Estimation}&FDI& \ding{51} & \ding{51} & Controllable False Alarms & Detection Delays & \cite{kurt2019secure} \\ \hline
PoW Consensus & \parbox[c][2.3em]{2.5cm}{\centering Data Tampering\\Unauthorized Access}  & \ding{55} & \ding{51} & Transparency& Complexity & \cite{mengelkamp2018blockchain} \\ \hline
Blockchain Signaling System & DDoS & \ding{55} & \ding{51} & Scalability & Complexity & \cite{rodrigues2019evaluating}\\ \hline
 PoW-PoS (trust-chain) & Insider (Sybil)& \ding{55} & \ding{51}& Resilient & Scalability & \cite{kolokotronis2019blockchain}\\ \hline
 Data Driven Trust Mechanism & Greyhole-Blackhole & \ding{51} & \ding{51} & Decentralization & Scalability & \cite{sivaganesan2021data}\\ \hline
 Permissioned Consortium (Ethereum) & FDI & \ding{51} & \ding{55} & High Accuracy & Complexity & \cite{ghiasi2021cyber}\\ \hline
\end{tabular}
} 
\label{table:blockchain}
\end{table*}

To understand the role of blockchain in improving smart grid security, it is important to emphasize the reliability it brings through its consensus mechanisms. Particularly, devices involved in power generation, transmission, distribution, or consumption initiate transactions by creating blocks. These blocks are validated by smart contracts, ensuring transactions adhere to a predefined set of rules. Consensus mechanisms like Proof of Work (PoW) and Proof of Stake (PoS) ensure all nodes accept the transaction \cite{zhang2019consensus}. Once consensus is achieved, the block is added to the blockchain, and the transaction is permanently recorded in the distributed ledger. This guarantees the immutability, security, and transparency of transactions. Figure \ref{figure:blockchain} illustrate the information flow between blockchain and smart grid operations, demonstrating the use of smart contracts for transaction validation, consensus mechanisms for network-wide agreement, and a distributed ledger to maintain the integrity and transparency of transaction data.


Blockchain technology provides a robust foundation for improving security in smart grids, particularly through the implementation of microgrids \cite{musleh2019blockchain, mollah2020blockchain}. These smaller, autonomous grids facilitate the integration and management of blockchain, which efficiently handles localized data and peer-to-peer energy trading. The decentralized nature of blockchain aligns well with microgrids' structure, improving their security, transparency, and resilience. It ensures that energy production and consumption data are securely recorded and easily accessible. Additionally, decentralization mitigates single points of failure and reduces vulnerability to cyber threats, significantly enhancing the resilience and robustness of the microgrid \cite{bani2018reliability}. Practical examples include the Brooklyn Microgrid project and the Power Ledger initiative in Australia, both of which demonstrate how local energy trading can be facilitated, making transactions easy and creating a more democratic energy market \cite{mengelkamp2018designing}. The TransActive Grid project further shows blockchain's potential to transparently log energy transactions, highlighting its applicability to smart grid technology \cite{orsini2019brooklyn}. Table \ref{table:blockchain} provides a comprehensive overview of blockchain theory, including detection and mitigation methods for various attacks, as well as their strengths and weaknesses.


{\bf Research challenges of blockchain techniques in smart grid security:} Although blockchain in smart grids brings numerous security benefits, it also present several research challenges \cite{koukaras2024integrating}. One major challenge is scalability. Current blockchain processing capabilities may not be sufficient to handle the high volume of transactions in a large-scale smart grid. The time required for increased transactions to be confirmed on the blockchain can introduce significant delays, reducing transaction speed, increasing costs, and inhibiting real-time energy trading  \cite{croman2016scaling, mengelkamp2018blockchain, monrat2019survey, agung2022blockchain}. Additionally, energy-intensive consensus mechanisms like Proof of Work are not sustainable for resource-constrained devices in smart grids \cite{vranken2017sustainability, mollah2020blockchain}. Research is needed to ensure low-latency and energy-efficient transaction validation and confirmation for real-time applications in smart grids. Interoperability is another critical issue. It is challenging to seamlessly integrate blockchain with current smart grid infrastructure and protocols  \cite{reyna2018blockchain, mollah2020blockchain}. The lack of standardization in blockchain technology can lead to compatibility issues between different systems and technologies. Ensuring interoperability is crucial for creating comprehensive solutions that enhance the security and efficiency of the grid ecosystem. Data privacy is also a significant concern. Ensuring data privacy while maintaining transparency on the blockchain is a complex problem, especially with sensitive data involved in smart grids. Disclosing transactional details can compromise user integrity and confidentiality, violating privacy preferences  \cite{mollah2020blockchain}. Research is needed to develop methods that ensure data privacy without compromising the transparency and security benefits of blockchain technology.


   \begin{table*}[t]
   \renewcommand\arraystretch{1.5}
   \caption{Summary of machine learning approaches for attack detection in Smart Grids}
				\centering
				\begin{tabular}{|m{3cm}<{\centering}|m{5cm}|m{7cm}|m{1.5cm}<{\centering}|}
					\hline
					\centering{\textbf{Learning Type}} & \centering{\textbf{Specific Methods}} & \centering{\textbf{Challenges}} & \textbf{References} \\
					\hline
					\centering{Supervised Learning} & Classification, Regression, Neural Network Methods & - Scarcity of high-quality labeled data \newline - Class imbalance and limited generalization to new attacks \newline - High computational intensity and black box nature of neural network models & \cite{ozay2015machine,azad2019transformation,yan2016detection,esmalifalak2014detecting,sakhnini2019smart,ahmed2018feature,zhang2011distributed,radoglou2018anomaly,boumkheld2016intrusion,yip2017detection,huang2021electricity,xue2019detection,li2018intrusion,yang2017improved,ford2014smart,ayad2018detection,niu2019dynamic,haghshenas2023temporal,takiddin2023generalized,musleh2019survey}\\
					\hline
					\centering{Unsupervised Learning} & Neural Network Methods, Clustering and Outlier Detection Methods, Other Detection Schemes & - High false positives/negatives and difficulty distinguishing between normal behaviors and attacks without ground truth \newline - Complexity due to dynamic nature of smart grids and high dimensionality of data & \cite{karimipour2019deep,wei2018false,chen2017outlier,meira2020performance,bhatia2019unsupervised,xu2018unsupervised,casas2012unsupervised,bhaumik2011clustering,pu2020hybrid,jiang2006clustering,syarif2012unsupervised,patcha2007overview,ahmed2019unsupervised,valdes2016anomaly} \\
					\hline
					\centering{Semi-supervised Learning} & Clustering-based Models, S3VM, Adversarial Autoencoders (AAEs), Generative-Adversarial-Based Semi-Supervised learning (GBSS), SS-Deep-ID, AE-GRU, GAN-RNN, Semi-WTC, ESFCM & - Curse of dimensionality and noise in the dataset \newline - Challenges in handling complex attack patterns, overfitting, and computational costs  \newline - Data variability and integrity issues, such as incompleteness or inconsistencies due to sensor or network failures & \cite{aamir2021clustering,haweliya2014network,ozay2015machine,zhang2020detecting,farajzadeh2021adversarial,abdel2021semi,dairi2023semi,li2022semi,rathore2018semi,qi2020semi,triguero2015self,zhang2020semi} \\
					\hline
					\centering{Ensemble Learning} & Bagging, Boosting, Stacking, Extremely Randomized Trees & - Computational complexity, especially in real-time systems \newline - Ensuring diversity among learners to prevent overfitting while maintaining accuracy & \cite{polikar2012ensemble,tama2021ensemble,aburomman2017survey,syarif2012application,ashrafuzzaman2020detecting,gaikwad2015intrusion,chen2018ensemble,cao2020novel,khoei2021ensemble,rashid2022tree,rajagopal2020stacking,acosta2020extremely,hazman2023lids,elgarhy2023clustering,starke2022cross,hu2020adaptive,sagi2018ensemble,kuncheva2014combining} \\
					\hline
					\centering{Reinforcement Learning} & SARSA, Q-Learning, DQN, Attention-aware DRL, Actor-Critic Methods, Deep Deterministic Policy Gradient (DDPG) & - Sparse and delayed rewards due to the infrequency of attacks \newline - Computational complexity vs. real-time decision-making \newline - Difficulty in creating flexible models for evolving grid dynamics and new attacks & \cite{sutton2018reinforcement,el2023electricity,kurt2018online,yan2016q,mnih2015human,an2019defending,li2022low,an2022data,huang2023attention,konda1999actor,feng2017deep,lillicrap2015continuous,abianeh2021vulnerability,yu2022fast} \\
					\hline
				\end{tabular}
				\label{table:ml}
			\end{table*}

\subsection{Machine Learning}
  

Machine learning (ML) focuses on developing algorithms and statistical models to identify patterns and make data-driven decisions. These patterns include correlations, trends, and anomalies within the data. By recognizing these patterns, ML algorithms can predict outcomes and detect irregularities. Through iterative processes, these algorithms improve their accuracy and performance over time \cite{jordan2015machine}. ML finds application in many fields, including network security \cite{khan2024adversarial}, natural language processing \cite{wireman2023comparison}, computer vision \cite{kappali2024computer}, agriculture \cite{liakos2018machine}, and medicine \cite{movahed2024tensor}.

         \begin{figure}[htbp]
			\centerline{\includegraphics[width=0.5\textwidth]{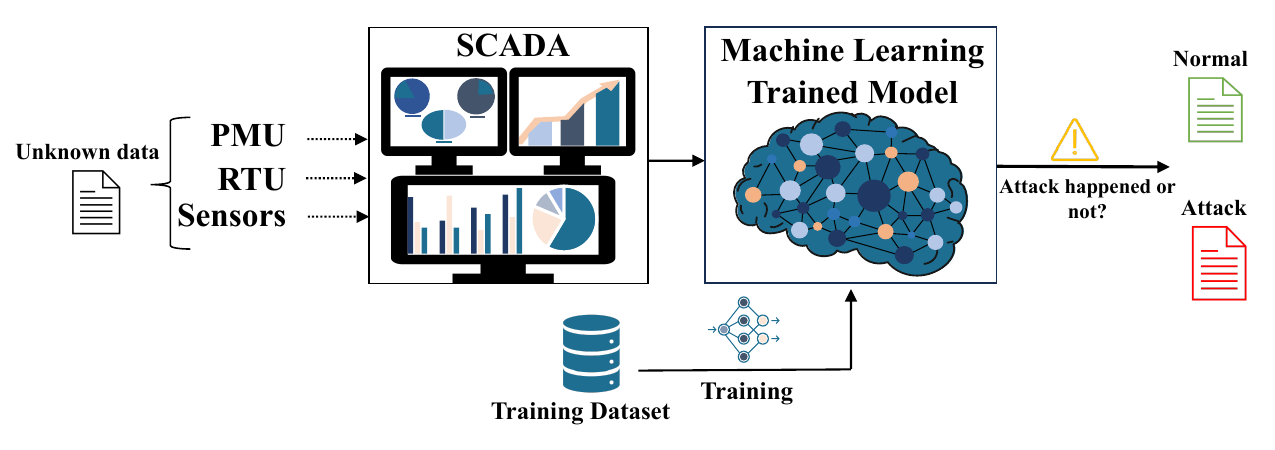}}
			\caption{Overview of Machine Learning-Based Attack Detection in Smart Grids.}
			\label{figure:ML}
        \end{figure}

ML has also been extensively applied into smart grids to improve its security. It assists smart grids in monitoring grid operations and detecting unusual activities that indicate security breaches. By analyzing vast amounts of data from different sources, such as PMUs, RTUs, and other sensors within the SCADA system, ML models can identify normal and suspicious activities. These models employ advanced algorithms and are trained on both regular and attack-pattern data, including power consumption, operational logs, and sensor readings, which allows them to effectively distinguish between normal and abnormal behaviors. Upon detecting a potential threat, an automated alert system immediately notifies grid operators, enabling them to take appropriate action to mitigate the risk. Figure \ref{figure:ML} provides a representation of this process.

ML can be deployed in either centralized or distributed configurations, depending on the specific infrastructure of the smart grid \cite{McMahan2016CommunicationEfficientLO,Lian2017AsynchronousDP,rashvand2024distributed,gholami2024digest}. In the centralized approach, data is aggregated and processed in a single location. This method simplifies data management and processing but can present scalability issues as data volumes increase. On the other hand, distributed ML processes data across multiple edge locations, which can improve model scalability and efficiency. However, it raises confidentiality concerns due to the need to transfer, store, and process data in various locations. To ensure data security in distributed settings, federated learning algorithms have been developed, allowing models to be trained across multiple decentralized devices without sharing raw data \cite{pmlr-v108-bagdasaryan20a,Mothukuri2021ASO,gholami2023fast,gholami2024improved,gholami2023federated,Tolpegin2020DataPA}. This approach helps mitigate the risks associated with data breaches and ensures that sensitive information remains protected while still benefiting from the advantages of distributed ML.

In this section, we examine various ML techniques that show promise in the field of smart grid security. We classify them into five categories: supervised, unsupervised, semi-supervised, ensemble, and reinforcement learning. Each category is discussed in detail, presenting its characteristics and limitations. We also provide recent examples to demonstrate their potential in enhancing smart grid security. Our studies are summarized in Table \ref{table:ml}.

\subsubsection{Supervised Learning Algorithms}
Supervised learning algorithms leverage labeled training data to develop models for attack detection and mitigation. They can be classified into three groups: classification, regression, and neural network methods. Each group has a distinct set of algorithms with specific application methods.
        
\underline{Classification Methods:} In supervised learning, a variety of algorithms have been applied to binary classification problems in the context of attack detection. 

The Perceptron operates as a fundamental learning algorithm using a weight vector and input samples to predict the occurrence of an attack. It iteratively adjusts the weights until a specific criterion is met. However, it only ensures convergence when the samples are linearly separable, making it most suitable for attack scenarios where measurements can be separated by a hyperplane \cite{ozay2015machine}. 

The K-Nearest Neighbor (KNN) algorithm is used to classify instances by identifying the most prevalent class among its k closest neighbors within the feature space. It has proven effective in detecting FDI attacks, particularly in smaller-scale smart grid systems \cite{azad2019transformation}\cite{yan2016detection}. However, when the feature vector size greatly exceeds the instance size, a situation known as the curse of dimensionality arises. We can adopt strategies such as feature selection algorithms, kernel methods like Support Vector Machines (SVMs), or reducing instance sizes to help mitigate the issue.

SVMs are designed to separate benign and corrupted measurements by computing a hyperplane in a transformed feature space. They introduce slack variables for flexibility when the samples are not linearly separable and prove their efficacy in high-dimensional spaces \cite{esmalifalak2014detecting,sakhnini2019smart,ahmed2018feature,zhang2011distributed}. 

Sparse Logistic Regression employs the Alternating Direction Method of Multipliers (ADMM) to solve classification problems, effectively reducing computational complexity while maintaining accuracy in large-scale or high-dimensional datasets. Additionally, methods such as Decision Trees (DT) and the Naive Bayes Classifier (NBC) have proven successful in the development of Intrusion Detection Systems (IDS) for smart grids. Decision Trees are adept at distinguishing between normal and malicious activities, providing high accuracy and true positive rates. Naive Bayes Classifiers, on the other hand, are particularly effective in detecting Denial of Service (DoS) attacks, including blackhole attacks, ensuring robust security for smart grid systems \cite{radoglou2018anomaly,boumkheld2016intrusion}.

A recent study by He et al. \cite{he2024detection} demonstrates the use of classification techniques in smart grids for detecting FDI attacks. They improve their model performance by using a cross-entropy loss function and adjusting the learning rate with Adam and cosine annealing algorithms.

\underline{Regression Methods:}
Regression techniques, such as linear regression, offer an alternative approach to classification methods for detecting smart grid attacks. For instance, Yip et al. presented two linear regression-based algorithms to identify defective and compromised smart meters in a neighborhood area network. These algorithms analyze reported energy consumption data to detect anomalies, enabling the successful identification of fraudulent consumers and faulty smart meters, and providing an accurate estimation of energy theft/loss \cite{yip2017detection}. However, it is important to note that while linear regression can handle large datasets and establish a quantitative relationship between variables, it requires a linear relationship between variables and is sensitive to outliers.
				
\underline{Neural Network Methods:}
Neural Networks (NN) are powerful machine learning models capable of identifying complex patterns within large datasets. In smart grids, NNs have shown great promise in detecting malicious activities, fraudulent behaviors, and anomalies.

Deep Neural Networks (DNN), a subset of NNs with multiple hidden layers, are particularly effective in learning intricate patterns. Their ability to learn complex representations makes them highly suitable for identifying various types of attacks, such as energy fraud in consumer energy consumption \cite{ford2014smart}. For example, autoencoders are effective in identifying anomalies within the high-dimensional and noisy data of smart grids. Huang et al. introduced a Stacked Sparse Denoising Autoencoder (SSDAE) for electricity theft detection \cite{huang2021electricity}. This model leverages sparsity, noise reduction, and the particle swarm optimization algorithm to extract robust features and set optimal error thresholds using a receiver operating characteristic curve. Additionally, the Extreme Learning Machine (ELM), a feed-forward NN, has been successfully employed to detect False Data Injection (FDI) attacks \cite{xue2019detection,li2018intrusion,yang2017improved}.

Recurrent Neural Networks (RNN) are designed to recognize patterns in sequences of data by utilizing their internal state (memory) to process variable-length sequences. They have been used to detect False Data Injection (FDI) attacks targeting the smart grid \cite{ayad2018detection}. Additionally, deep learning frameworks that incorporate both Convolutional Neural Networks (CNN) and Long Short-Term Memory (LSTM) networks have been proposed to detect anomalies caused by FDI attacks \cite{niu2019dynamic}. Another approach is the Temporal Graph Neural Network (TGNN) proposed by Haghshenas et al. \cite{haghshenas2023temporal}, which detects and localizes FDI attacks in smart grids by leveraging GNN to capture system topological information and state measurements. Furthermore, Takiddin et al. \cite{takiddin2023generalized} proposed a strategy using a Generalized Graph Neural Network and a graph autoencoder (GAE) to detect FDI attacks. This method captures spatio-temporal features to improve generalization and effectiveness against unseen attacks.
    
{\bf Research challenges of supervised learning in smart grid security:}
Despite their effectiveness, supervised learning methods in smart grid security face several challenges, including the need for extensive labeled data, handling data imbalance due to the rarity of attack events, and ensuring scalability to process vast amounts of real-time data. The scarcity of high-quality labeled data limits model training effectiveness and impedes generalization to new attacks. This is further exacerbated by class imbalance, where normal data far outnumbers attack samples \cite{musleh2019survey}. Additionally, these methods often suffer from high computational intensity, significant data requirements for training, and a lack of interpretability. Further, models must be adaptable to evolving cyber threats, protect privacy while using sensitive data, and manage substantial computational resource demands.

\subsubsection{Unsupervised Learning Algorithms}

Unlike supervised learning that relies on labels, unsupervised learning leverages ML algorithms to analyze and cluster unlabeled datasets. These algorithms discover hidden patterns without the need for human intervention. Given the computational challenges and high false alarm rates of supervised learning in large-scale anomaly detection, unsupervised learning emerges as a more effective alternative for identifying anomalies without the need for labeled data \cite{chandola2009anomaly}.

\underline{Neural Network Methods:}
In unsupervised learning, Deep Belief Networks (DBN) and autoencoders have been particularly effective to detect anomalies and attacks in smart grids. In \cite{karimipour2019deep, wei2018false}, DBNs are utilized, leveraging unsupervised learning for initial weight allocation and subsequent model parameter fine-tuning, thereby outperforming SVM-based detection methods. Autoencoder, another powerful tool in unsupervised anomaly detection, have been employed in ensemble methods with varying structures and connection densities to enhance efficiency, diversity, and training time \cite{chen2017outlier}. Further, \cite{meira2020performance} incorporated autoencoders into six one-class algorithms for anomaly detection. Autoencoder-based classifiers have demonstrated notable effectiveness in real-time detection of various DDoS attacks by utilizing the predictability of TCP traffic, outperforming supervised methods in detecting new attacks \cite{bhatia2019unsupervised}. Additionally, \cite{xu2018unsupervised} introduces Donut, an innovative unsupervised anomaly detection algorithm based on Variational Autoencoder (VAE) for Key Performance Indicators (KPIs), which are measurable values that demonstrate how effectively an organization is achieving key objectives. Although originally designed for web applications, this algorithm can be adapted to smart grid security by monitoring the grid's operational data in terms of power quality, load forecasts, outage durations, energy production levels, distribution efficiency, incident response times, and cybersecurity incidents.
				
\underline{Clustering and Outlier Detection Methods:}
Clustering and outlier detection techniques in unsupervised learning also hold significant potential for robust anomaly detection. For example, \cite{casas2012unsupervised} introduces UNIDS, a system that employs Sub-Space Clustering for outlier detection, demonstrating the power of clustering-based approaches in unsupervised learning. \cite{bhaumik2011clustering} further explores this field by utilizing the K-Means clustering algorithm to identify suspicious activities within networks.
Several hybrid clustering techniques have been proposed, including the method presented by \cite{pu2020hybrid}, which combines Sub-Space Clustering (SSC) and One Class SVM (OCSVM) for unsupervised anomaly detection. This approach, tested on the NSL-KDD dataset, detects attacks without prior knowledge. Furthermore, Jiang et al. \cite{jiang2006clustering} introduced a novel method to determine the outlier factor of clusters, measuring the deviation degree of a cluster and computing the cluster radius threshold. Their improved nearest neighbor (INN) method and clustering-based unsupervised intrusion detection show higher accuracy in detecting unknown intrusions. Syarif et al. \cite{syarif2012unsupervised} evaluated the performance of various clustering algorithms in network anomaly detection. They found that while traditional misuse detection techniques struggled with high unknown intrusion rates, clustering-based anomaly detection showed promising accuracy. Finally, \cite{patcha2007overview} provides a comprehensive overview of various clustering and outlier detection schemes, illustrating their integral role for effective intrusion detection in smart grids.

\underline{Other Detection Schemes:} Isolation forest, adaptive resonance theory, and self-organizing maps have also emerged as powerful unsupervised learning tools for anomaly detection. In \cite{ahmed2019unsupervised}, the isolation forest technique is used to detect Covert Data Integrity Assaults (CDIA) in smart grid networks. It isolates anomalies using a binary tree structure, based on the premise that anomalies are fewer and require fewer splits to isolate. \cite{valdes2016anomaly} introduces a machine learning-based IDS for smart grids, using Adaptive Resonance Theory and Self-Organizing Maps to identify cyber attacks, assess their impact on physical measurements, and differentiate between normal, faulty, and malicious states.

{\bf Research challenges of unsupervised learning in smart grid security:} Unsupervised learning for smart grid attack detection faces several key challenges. First, it is inherently difficult to distinguish between normal behavior and cyber-attacks without labeled data. Additionally, it is challenging to measure the performance and ensure the reliability of unsupervised models in the absence of a ground truth. Collecting high-quality data is essential to develop effective unsupervised learning models that can accurately detect anomalies and cyber attacks. Furthermore, smart grids produce vast amounts of data continuously, and unsupervised learning models must be capable of handling this volume in real-time to provide timely detection and response to potential threats. The evolving nature of cyber threats also require the models to be regularly updated to recognize new attack patterns and strategies. Moreover, unsupervised models often operate as black boxes, providing little insight into how they reach their conclusions. This lack of interpretability in detection decisions hampers trust and deployment, as stakeholders need to understand and trust the system’s decisions.

\subsubsection{Semi-supervised Learning Algorithms}
Semi-supervised learning models effectively utilize both labeled and unlabeled data, offering robust solutions for various cybersecurity challenges in smart grids. \cite{aamir2021clustering} demonstrates that the clustering-based semi-supervised models are particularly effective for DDoS attack detection. Meanwhile, the semi-supervised SVM model, presented in \cite{haweliya2014network} and \cite{ozay2015machine}, handles network intrusion detection and smart grid attack detection with an impressive accuracy of up to 90 percent. Leveraging the strengths of autoencoders and GANs, Adversarial Autoencoders (AAEs) proposed by \cite{zhang2020detecting} provide an effective solution for detecting unobservable FDI attacks in smart grids. Similarly, the Generative-Adversarial-Based Semi-Supervised learning (GBSS) framework, discussed in \cite{farajzadeh2021adversarial}, utilizes conditional GANs for diagnosing cyber attacks and faults in power grids, outperforming other semi-supervised models under challenging learning conditions. 

The SS-Deep-ID model, proposed by \cite{abdel2021semi}, incorporates a multiscale residual temporal convolutional module and a traffic attention mechanism. It proves superior in various metrics and demonstrates computational effectiveness in real-time intrusion detection for smart grids. Dairi et al. present two innovative methods for anomaly detection: AE-GRU, which combines a GRU-based stacked autoencoder, and GAN-RNN, which integrates a GAN with an RNN for enhanced cyber attack detection in smart grids \cite{dairi2023semi}. The Semi-WTC framework introduces an innovative semi-supervised method that enhances attack detection in cybersecurity by combining weight-task consistency and a recurrent prototype module \cite{li2022semi}. 

Rathore et al. introduce a semi-supervised learning-based distributed attack detection framework for smart grids, utilizing a fog-based approach to enhance security and efficiency through the novel ESFCM method \cite{rathore2018semi}. Another study utilizes semi-supervised learning combined with deep feature extraction techniques to detect cyber-attacks by leveraging normal operation data from PMUs, enabling the identification of anomalies without requiring extensive examples of attack patterns \cite{qi2020semi}.

{\bf Research challenges of semi-supervised learning in smart grid security:} Semi-supervised learning models face challenges in handling complex attack patterns, overfitting, and the computational costs associated with large, high-dimensional smart grid data. Reducing data complexity while retaining critical features is challenging because standard ML methods are not designed to effectively handle both labeled and unlabeled data simultaneously. This necessitates the development of new strategies to manage this combination effectively \cite{triguero2015self}. The variability, incompleteness, and inconsistency of smart grid data, caused by sensor malfunctions or network issues, further complicate the learning process. Moreover, the dynamic evolution of attack vectors necessitates continuous adaptation of semi-supervised learning models. This adaptation is constrained by the scarcity of labeled data for new threats, making it challenging for these models to effectively detect emerging attacks. Finally, the computational costs associated with processing large volumes of smart grid data and the need for real-time anomaly detection increase the complexity, necessitating robust, efficient, and scalable solutions.
			
\subsubsection{Ensemble Learning}

Ensemble learning combines multiple classifiers into a meta-classifier to enhance detection accuracy and robustness \cite{polikar2012ensemble}. It has proven their effectiveness across various applications in smart grid security. For example, \cite{ashrafuzzaman2020detecting} demonstrates the efficacy of ensemble learning for stealthy FDI attack detection in smart grids by combining both supervised and unsupervised classifiers. Elgarhy et al. develop a strategy to secure electricity theft detectors against evasion attacks by clustering smart grid consumers based on their consumption patterns. They assign specific detectors to each cluster and reinforce them with an ensemble of models \cite{elgarhy2023clustering}. 

Ensemble ML techniques, including bagging \cite{breiman1996bagging}, boosting \cite{bartlett1998boosting}, stacking \cite{wolpert1992stacked}, and extremely randomized trees, can significantly improve the effectiveness of attack detection in IDS for smart grids \cite{tama2021ensemble, aburomman2017survey, syarif2012application}.

Bagging, or bootstrap aggregating, trains multiple base learners on different subsets of the original dataset, providing diversity among the learners and reducing model variance. When implemented using REPTree as the base class, it exhibits superior performance in terms of classification accuracy, lower false positives, and reduced model-building time in IDSs \cite{gaikwad2015intrusion}. Moreover, \cite{chen2018ensemble} demonstrates that both bagging and adaptive boosting can improve the reliability of cyber attack detection in power systems.

Boosting constructs a robust classifier from multiple weak ones by iteratively training on subsets of the total dataset and assigning increased weight to misclassified examples, thus concentrating the model on challenging cases. The CKS-FCS-FLGB method has utilized boosting to effectively detect varied FDI attacks even in small-sample datasets, enhancing the stability and resistance of smart grid systems \cite{cao2020novel}. Hazman et al. propose IDS-SIoEL, a sophisticated IDS for cyber-physical systems that employs AdaBoost and feature selection to enhance security performance \cite{hazman2023lids}.

Stacking combines the predictions of various base learning models via a meta-learner to improve detection probability, reduce false alarm rates, minimize miss detection rates, and enhance overall accuracy \cite{khoei2021ensemble}. \cite{rashid2022tree} demonstrated the efficacy of a tree-based stacking ensemble technique for IDSs, while \cite{rajagopal2020stacking} underscored the value of stacking for effective network intrusion detection using heterogeneous datasets.

The Extremely Randomized Trees method introduces a level of randomness and robustness by randomizing the selection of features and thresholds for each split in the tree. When coupled with Kernel Principal Component Analysis for dimensionality reduction, it displayed superior efficiency and accuracy in detecting stealthy cyber attacks in smart grid networks \cite{acosta2020extremely}.
			
{\bf Research challenges of ensemble Learning in smart grid security:} Detecting attacks in smart grids using ensemble learning faces several significant challenges. One major issue is handling heterogeneous and multi-sourced data, including both cyber and cyber-physical systems, which increases the complexity of identifying attacks \cite{starke2022cross, hu2020adaptive}. Another challenge is the computational complexity of combining multiple machine learning models, especially in real-time systems where rapid decision-making is crucial \cite{sagi2018ensemble}. Additionally, ensuring diversity among individual learners is critical; too much similarity can lead to overfitting, while too much variation may reduce accuracy \cite{kuncheva2014combining}. These challenges complicate the development and deployment of effective ensemble learning models for smart grid security.

\subsubsection{Reinforcement Learning}

\begin{figure}[htbp]
				\centering
				\includegraphics[width=0.5\textwidth]{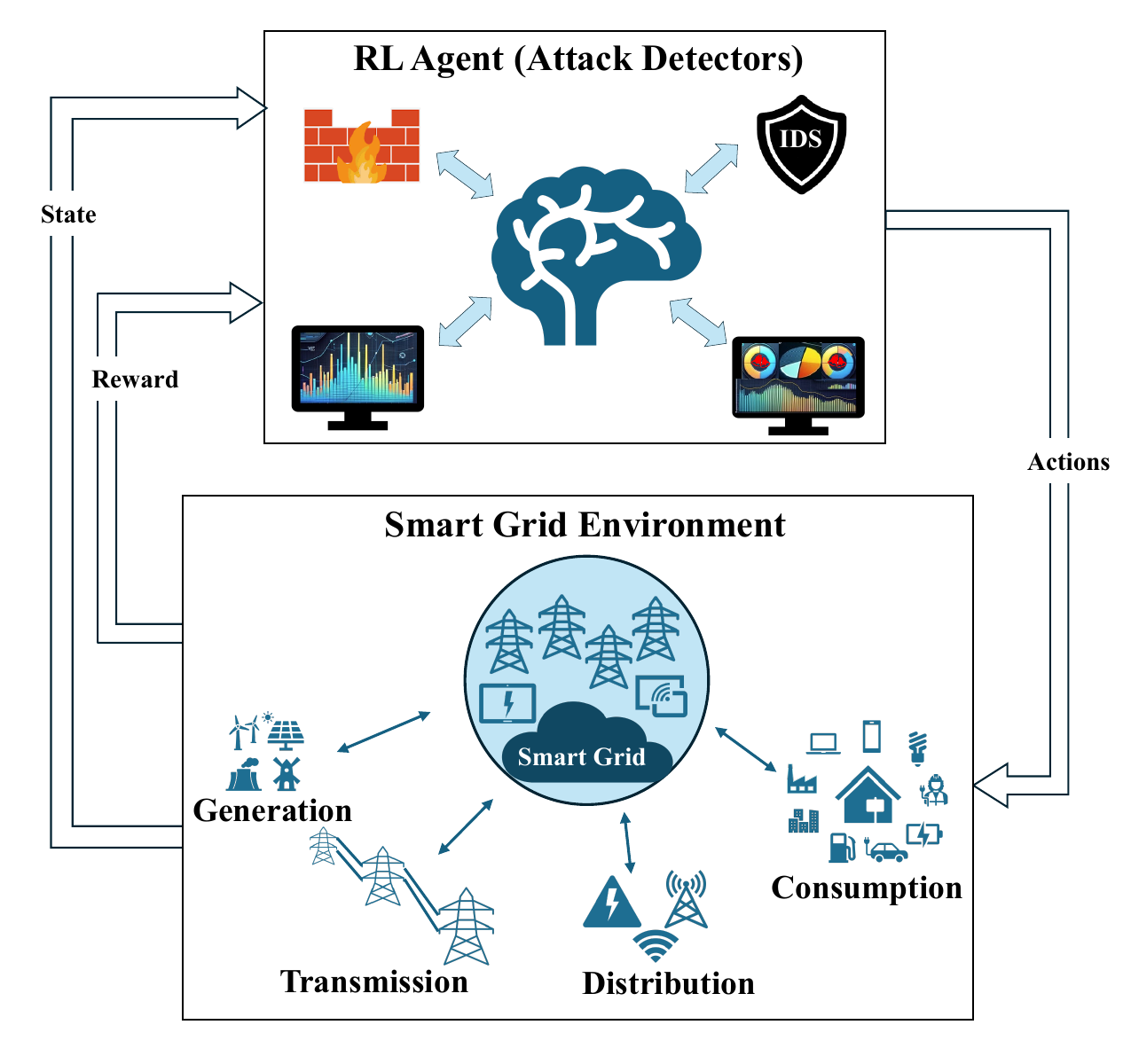}
				\caption{A representation of the Reinforcement Learning process in Smart Grids for attack detection.}
				\label{figure:RL}
			\end{figure}

Reinforcement Learning (RL) trains algorithms to learn from their environments. In a smart grid, RL works through an iterative interaction process between an agent (e.g., a detection system) and the environment (e.g., the smart grid infrastructure). The agent observes the current state of the environment, such as network traffic patterns indicating potential unauthorized access, power consumption readings revealing unusual spikes suggestive of electrical theft or equipment failure, and the operational status of grid components showing signs of malfunction or tampering. The agent performs actions within the environment to achieve specific goals, such as detecting anomalies or security breaches. Following each action, the agent receives feedback in the form of rewards to learn and improve its strategy. The agent policy, which maps observed environmental states to actions, evolves to maximize cumulative rewards, enabling more accurate and quick detection of attacks \cite{sutton2018reinforcement}. Figure \ref{figure:RL} illustrates the RL framework, depicting the agent as a detector, the environment as the smart grid infrastructure, and the crucial interaction driven by the state of the system.

RL methods such as State Action Rewarded State Action (SARSA), Q-Learning, Deep Q-Network (DQN), Attention-aware deep reinforcement learning (Attention-aware DRL), Actor-Critic Methods, and Deep Deterministic Policy Gradient (DDPG) have been extensively applied to address various security challenges in smart grids. 

State Action Rewarded State Action (SARSA) is a model-free RL method that learns the Q-value for state-action pairs to guide the policy \cite{sutton1998introduction}. It enables robust, proactive attack detection without requiring specific attack models and responds sensitively to minor deviations from normal operations, thus limiting the attacker's action space. This approach has been successfully implemented for online cyber attack detection in smart grids \cite{kurt2018online}.

Q-Learning is another model-free algorithm that learns the value of an action in a particular state \cite{watkins1992q}. In smart grids, it has been applied to analyze vulnerabilities against sequential topology attacks \cite{yan2016q}, successfully identifying critical attack sequences. However, it struggles with large and continuous state-action spaces, and can hardly identify the best actions when faced with delayed rewards.

Deep Q-Networks (DQNs) were introduced to overcome the limitations of Q-Learning by combining Q-Learning with deep neural networks \cite{mnih2015human}. DQNs have been employed to detect data integrity attacks in AC power systems, offering improved efficiency by avoiding the curse of dimensionality \cite{an2019defending}. Li et al. further enhanced this approach with a DQN-based DRL algorithm that achieves both high accuracy and low latency in detecting cyber attacks in smart grids \cite{li2022low}. This method incorporates a dynamic AC system model, a continuous state-space DQN within a Markov Decision Process (MDP) framework, and a unique reward function to balance detection delay and accuracy. El-Toukhy et al. introduce a method to detect electricity theft attacks in smart power grids using DQN and Double DQN (DDQN). These techniques adapt to dynamic theft behaviors by learning optimal actions through exploration and exploitation mechanisms \cite{el2023electricity}. Other researchers further enhanced the DQN methodology by integrating it with LSTM networks, effectively addressing the continuously changing nature of grid states and the rare occurrence of attack states \cite{an2022data}. Despite these improvements, DQNs still face challenges with overestimating Q-values and maintaining stability during the learning process.

Attention-aware DRL integrates an attention mechanism into the DRL framework, allowing it to selectively focus on critical parts of the state for feature extraction while avoiding distractions from irrelevant details. This approach improves the representation and distinguishability of states and has demonstrated effectiveness in detecting FDI attacks in smart grids \cite{huang2023attention}.

There are also some DRL-based methods called Actor-Critic methods balance the benefits of value-based and policy-based methods by having two separate components, an actor that decides the policy and a critic that evaluates it \cite{konda1999actor}. An application of these methods in a real-world context can be seen in the work by Feng and Xu \cite{feng2017deep}, which presents an optimal online defense strategy for CPS under unknown cyber-attacks, employing a novel cyber-state dynamics model and a game-theoretical actor-critic NN structure, further enhanced by a DRL algorithm, leading to real-time, accurate, and timely learning and implementation of optimal defense and worst attack policies. 

Actor-Critic methods balance the benefits of value-based and policy-based methods by utilizing two separate components: an actor that decides the policy and a critic that evaluates it \cite{konda1999actor}. Feng and Xu present an optimal online defense strategy that employs a novel cyber-state dynamics model and an actor-critic neural network for unknown attacks mitigation in cyber-physical systems \cite{feng2017deep}. This approach enables real-time, accurate, and timely learning and implementation of optimal defense and worst attack policies.

Deep Deterministic Policy Gradient (DDPG) is an algorithm that combines DQN and actor-critic approaches, specifically designed to handle environments with continuous action spaces \cite{lillicrap2015continuous}. It has been successfully applied to detect FDI cyber-attacks and identify vulnerabilities in conventional index-based cyber-attack detection systems for DC microgrids \cite{abianeh2021vulnerability}.

{\bf Research challenges of reinforcement Learning in smart grid security:}
Although RL methods offer many opportunities to enhance smart grid security, several challenges remain to be addressed \cite{wang2024reinforcement, sharif2024smart}. First, high-quality data is essential for training effective RL models, but obtaining sufficient labeled data, especially for rare cyber-attack events, is difficult. The infrequency of attacks leads to sparse and delayed rewards, complicating the development of effective response policies. Additionally, RL models need to balance the exploration of new strategies with the exploitation of known ones; excessive exploration can lead to suboptimal performance and compromise grid security. Real-time processing capabilities are also required for prompt decision-making, but this is challenging due to the computational demands of many RL algorithms. Furthermore, the dynamic nature of cyber threats requires RL algorithms to continually and rapidly adapt to new and evolving attack patterns. However, the complexity of smart grid systems and the variability of attacks make it difficult to create precise RL models that can accommodate evolving grid dynamics and new threats. 
		
\section{Future research directions}
\label{section:future}

We have conducted a comprehensive overview of various emerging detection and mitigation strategies in smart grids, discussing their advantages and research challenges. Despite notable advancements, these challenges highlight the critical need for continued research and development to enhance the effectiveness and reliability of these techniques and ensure the resilience and integrity of future smart grid infrastructures. In this section, we present potential research directions that are crucial for advancing these detection and mitigation techniques. 

\subsection{Research Directions for Emerging Techniques}

\underline{Research directions for game theory methods:} Game theory has shown great potential for developing decision-making, analysis, and control algorithms in smart grid security. Future research is needed to refine the proposed models, develop practical algorithms, and decentralize decision-making process \cite{alpcan2003game}. Research could focus on developing game-theoretic models that consider partial network failures instead of complete blackouts, and incorporating varying levels of defense success to create more realistic and adaptable security models. Additionally, it is crucial to account for the criticality of different nodes in the network, concentrating defense efforts on the most important nodes to optimize resource allocation. Research could further expand the scale of existing models to include nonlinear relationships, which will better represent grid dynamics and account for performance drops due to faulty nodes \cite{shan2020game}. Also, probabilities of natural failures and cyber attacks should be independently considered to make models more comprehensive and realistic. Furthermore, future research should aim to decentralize the decision-making processes to improve the resilience and efficiency of smart grid operations.

\underline{Research directions for graph-based methods:} Graph-based methods have shown promise in enhancing smart grid security by modeling and analyzing complex relationships and interactions within the grid. Future research could focus on several key areas to further advance these techniques. First, It is crucial to address scalability issues and optimize computational efficiency to deploy graph-based methods in large-scale smart grid systems. One promising direction is to reduce the complexity of the attack graph by abstracting multiple vulnerabilities on a host with similar effects into a generic vulnerability \cite{ammann2002scalable}.
Another direction is to refine algorithms for traversing attack graphs to identify the most cost-effective network configuration changes, thereby protecting critical resources \cite{wang2006minimum}. Additionally, research could expand graph-based models to handle dynamic changes in the smart grid, such as fluctuating power demands and varying network configurations, to improve real-time analysis and response capabilities. It is also essential to develop more sophisticated algorithms to detect and mitigate cyber attacks by leveraging graph theory ability to represent network topologies and dependencies. Graph-based approaches could further incorporate machine learning and artificial intelligence to enhance predictive capabilities, enabling the early identification of potential threats.
 
\underline{Research directions for blockchain methods:} Blockchain methods hold significant potential to enhance smart grid security by providing decentralized, transparent, and tamper-proof transaction records. Future research could focus on several areas to advance these techniques. Firstly, it is essential to develop scalable blockchain architectures tailored to the high transaction volumes and real-time requirements of smart grids. Improving consensus algorithms is also critical to enhance transaction speed and reduce energy consumption. Additionally, research should explore the interoperability of blockchain systems with existing smart grid infrastructure to ensure seamless integration and operation. Further, it is important to investigate the economic implications and cost-effectiveness of implementing blockchain for widespread adoption in smart grids. Specifically, a technological evaluation of blockchains should be conducted to assess their scalability, robustness, resource usage, and transaction costs, determining the practical feasibility and efficiency of blockchain implementations in microgrid energy markets \cite{mengelkamp2018designing}. Addressing these research directions will help realize the full potential of blockchain methods in securing smart grid operations.

\underline{Research directions for machine learning methods:} Machine learning methods offer promising advancements for smart grid security, yet several key areas require further research to fully realize their potential. Current applications in smart grids face limitations such as scalability, data scarcity and quality, model interpretability, computational complexity, and adaptability, privacy concerns. Future research should focus on developing more efficient and scalable machine learning algorithms capable of handling the vast amounts of data generated by smart grids. Advanced data preprocessing and augmentation techniques should be developed to enhance the quality of training data, addressing challenges of sparse and incomplete data. Collaborative data sharing frameworks among different smart grid operators could further enhance data availability, providing a more comprehensive dataset for model training and validation. Additionally, research could aim to develop models that provide clear explanations for their predictions and decisions, facilitating their integration into existing smart grid infrastructures. It is also crucial to reduce the computational complexity of machine learning algorithms to ensure they can be deployed on resource-constrained devices in smart grids. Real-time processing capabilities will be essential for machine learning to continuously learn from new data and adapt their strategies accordingly. Furthermore, research should address privacy concerns by incorporating advanced privacy-preserving techniques, such as federated learning and differential privacy, to protect sensitive data while enabling effective machine learning model training.

We will also explore new and emerging ML techniques and concerns to enhance smart grid security. One promising direction is the application of large language models (LLMs). LLMs have the potential to understand complex attack patterns and detect sophisticated zero-day attacks. Incorporating LLMs into smart grid systems may significantly improve anomaly detection. Another critical area is addressing the increasing concern of adversarial machine learning attacks. As ML algorithms become more prevalent in smart grids, adversarial attacks pose a significant threat by misleading existing models and bypassing detectors. Addressing these threats is crucial to maintain the integrity of smart grid security systems.

In what follows, we will explore the potential and challenges of using LLMs in smart grids and examine adversarial ML attacks, their mechanisms, and mitigation strategies.




\subsection{Large Language Models}

LLMs like OpenAI's GPT \cite{achiam2023gpt} and Google's Gemini \cite{reid2024gemini} hold significant promise in the field of cybersecurity. They offer substantial potential for improving the cybersecurity of smart grids, particularly in digital substations \cite{zaboli2023chatgpt}. For instance, LLMs are effective for detecting DDOS attacks when used with few-shot learning or fine-tuning \cite{guastalla2023application}. Unlike traditional ML models, which require frequent re-training to accommodate new attack patterns, LLMs can interpret context and respond to novel threats without explicit re-training. By leveraging LLMs, we can analyze vast datasets and system logs to identify patterns and detect irregularities that indicate potential cyber-attacks \cite{gupta2023chatgpt}. For example, LLMs can dynamically understand and adapt to new threats, offering a robust solution for anomaly detection in IEC 61850-based communications, such as generic object-oriented system events and sampled values messages \cite{hussain2023novel}.

However, deploying LLMs in smart grid security comes with several limitations that need to be addressed. One major concern is their tendency to memorize specific sequences from their training data, which can raise privacy and security issues \cite{biderman2024emergent}. This memorization can inadvertently lead to the reproduction of personal information, highlighting the need for meticulous data handling and privacy measures. Furthermore, while LLMs show promise in understanding complex attack patterns and data, their deployment for automatic vulnerability detection remains challenging due to their limitations in action-oriented capabilities. The ReAct, short for Reasoning and Acting, aims to bridge this gap by enabling LLMs to dynamically reason about their actions, update plans based on environmental feedback, and improve task performance. Integrating reasoning and acting makes LLMs more adaptive and capable of handling dynamic and evolving scenarios, which is crucial for tasks (e.g., automatic vulnerability detection in smart grids) that require both understanding and proactive interventions \cite{yao2022react}. Additionally, LLMs are susceptible to targeted attacks such as bad data injection and domain knowledge extraction, emphasizing the need for robust threat models and validation mechanisms \cite{li2024risks}. Moreover, pre-trained LLMs are not ready for deployment as-is; they need to be fine-tuned or used with few-shot learning to provide context-specific solutions \cite{brown2020language}. 

These challenges highlight the need for comprehensive strategies to ensure the secure deployment of LLMs in smart grids. By carefully addressing these concerns, LLMs could become a promising solution, potentially replacing traditional machine learning methods and aiding in the development of advanced IDSs, thereby enhancing overall system resilience. 


\subsection{Adversarial Machine Learning}

Adversarial machine learning attacks exploit vulnerabilities in ML models, leading to incorrect predictions and compromising the reliability and accuracy of decision-making processes \cite{goodfellow2014explaining}. They aim to create malicious inputs, known as adversarial examples, to manipulate ML models, posing a significant risk to the reliability of ML based detection and mitigation techniques. In particular, the adversarial examples can blur the line between grid vulnerabilities and normal grid performance, creating an attack surface for adversaries to exploit breaches in smart grids without being detected. 

\begin{figure}[htbp]
\centerline{\includegraphics[width=3.5in]{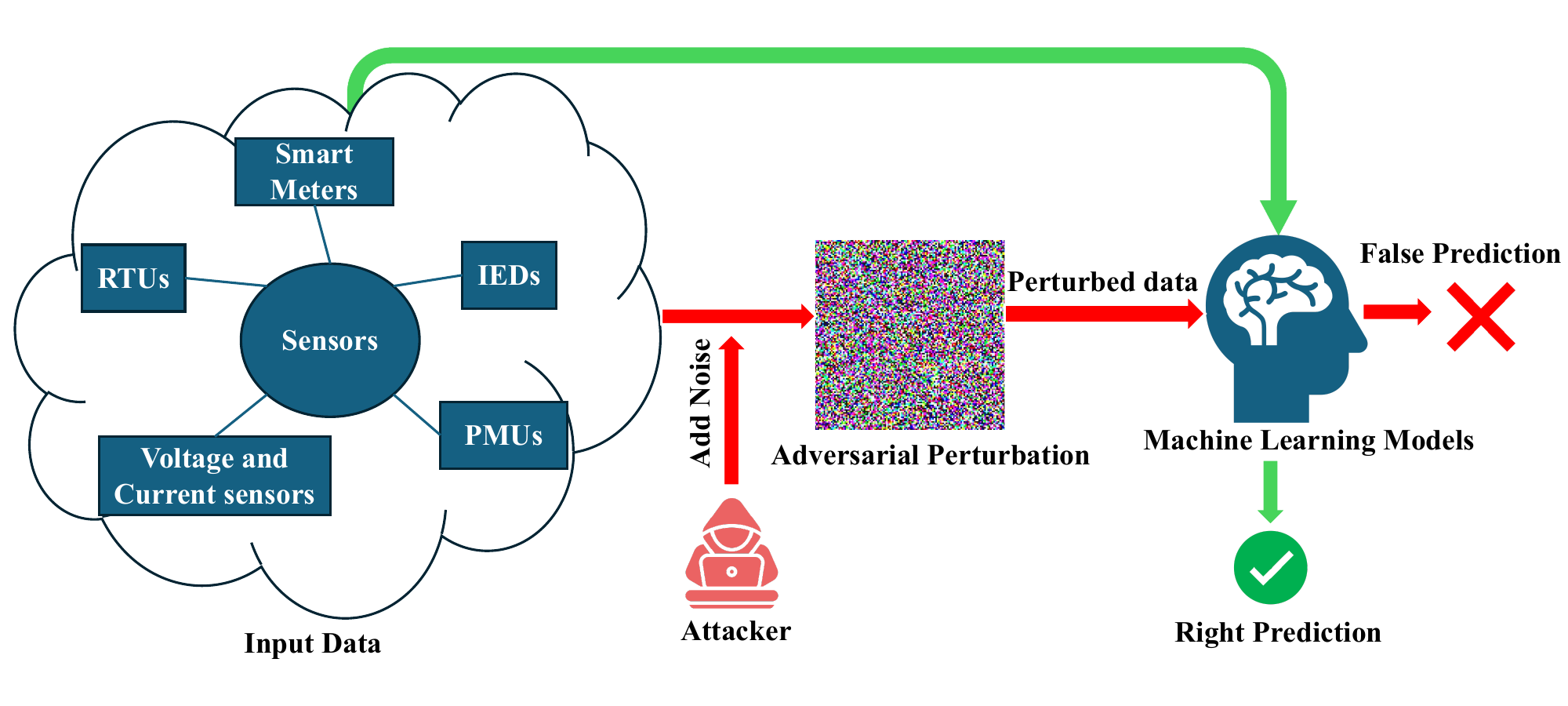}}
\caption{Illustration of an adversarial attack in a smart grid.}
\label{figure:adversarial}
\end{figure}

Adversarial machine learning attacks can be launched in two main ways: data poisoning attacks and evasion attacks. Poisoning attacks inject adversarial data into the training set to deceive the model during the training stage \cite{munoz2017towards, shafahi2018poison}. These attacks can alter the learning process of the model, leading to incorrect predictions even on clean data. Evasion attacks, on the other hand, occur during the inference stage. They make subtle changes to the test data, leading the model to misclassify inputs it has learned to recognize \cite{flowers2019evaluating, zhang2015adversarial, biggio2013evasion}. Evasion attacks are more prevalent as they are easier to execute and do not require access to the training phase. Attackers can carry out evasion attacks with just query access to the model, making them a more practical and immediate threat. Figure \ref{figure:adversarial} illustrates the process of an adversarial attack on ML models in smart grids.


Adversarial attacks can be categorized into three main types: white-box, gray-box, and black-box attacks. In white-box attacks, the attacker has complete knowledge of the target model, including its architecture, parameters, and training data. It allows the attacker to generate highly effective adversarial examples by exploiting specific vulnerabilities in the model \cite{zhou2024investigating, roshan2024untargeted,mahbub2024robustness}. In gray-box attacks, the adversary has partial knowledge of the target model, such as knowing the model architecture but not having access to internal weights or specific training data. Adversarial examples are often generated using a surrogate model that has a similar architecture to the target model, allowing the attacker to approximate the behavior of the target model and identify vulnerabilities \cite{jo2024exploring, apruzzese2022mitigating, wang2022gray}. In black-box attacks, the attacker has no direct knowledge about the internals of the target model. Instead, they rely on querying the model and observing its outputs to infer ways to generate adversarial examples \cite{papernot2017practical,suya2024sok,liu2024boosting,chen2017zoo,zhou2018transferable,dong2018boosting,thomas2022dynamic,ilyas2018black}. 


These attacks can severely disrupt smart grid operation by causing misclassifications in attack detection, leading to incorrect energy predictions and faulty grid management decisions, potentially resulting in power disruptions \cite{goodfellow2014explaining, he2017adversarial, hitaj2017deep, papernot2016limitations, carlini2017towards}. Therefore, it is crucial to develop techniques to enhance the resilience of ML models and make them robust against these attacks.


Defending against adversarial ML attacks requires a comprehensive approach. One effective strategy is adversarial training, where models are intentionally trained on adversarial examples to improve their robustness and resilience \cite{madry2017towards}. Defensive distillation, which trains a secondary model using the soft probabilities derived from a primary model, reduces the model sensitivity to small input changes \cite{papernot2016distillation}. Regularization methods, such as dropout and weight decay, can also enhance model robustness. Another strategy uses input transformations to move potential adversarial examples far from the decision boundary of the model, combined with detectors that identify inputs deviating from expected patterns \cite{carlini2017magnet}. Thermometer encoding offers a new way to encode categorical data. Instead of using a one-hot representation, it employs a cumulative representation where each binary bit corresponds to a temperature threshold, making it much harder for attackers to manipulate inputs without being detected \cite{buckman2018thermometer}. Further, game theory presents a promising approach to enhance robustness against adversarial attacks. Equilibrium strategies can create adaptive defenses that evolve in response to the adversary's tactics, making machine learning models more resilient in dynamic environments \cite{zhou2019survey, dasgupta2019survey}. Finally, continuous model evaluation and updating are essential to adapt to new attack strategies and maintain security.


It remains a challenging task to address adversarial machine learning attacks, as no single solution fits all purposes \cite{zuhlke2024adversarial, zhao2024evaluating}. Heuristic-based defense mechanisms cannot protect against all types of attacks. The transferability of adversarial examples is another major challenge; if an attack works on one model, it is likely to work on another model with a different architecture or training data, allowing attackers to refine their strategies on less secure models. In addition, while the ability to predict and prevent adversarial attacks is desirable, it usually requires high computational cost and may reduce the model accuracy on legitimate inputs. Further, defenses that harden a model against adversarial inputs often come with trade-offs, such as decreased accuracy on clean inputs or increased model complexity \cite{yuan2019adversarial}.These challenges against adversarial machine learning attacks underscore the need for ongoing research to develop new and flexible defense strategies, which will mitigate risks and enhance the resilience of machine learning models in smart grids \cite{wiyatno2019adversarial, yuan2019adversarial, ilahi2021challenges}. 






\section{Conclusion}
\label{section:conclusion}
 
In this study, we have explored the intricate nature of cyber, cyber-physical, and coordinated attacks on smart grids, offering a comprehensive analysis of various attack types and their implications for smart grid security. We critically examine and compare a range of detection and mitigation strategies, highlighting their potential benefits and limitations. Additionally, we discuss potential future research directions for emerging detection techniques, including the deployment of LLMs in smart grid security and the emerging threat of adversarial machine learning. Our study not only enhances the understanding of current threats but also identifies key areas for future research, emphasizing the urgent need for advanced, robust detection algorithms to keep pace with the rapidly evolving threats to smart grids.

\bibliographystyle{IEEEtran}
\bibliography{ref}
\end{document}